\documentclass{LMCS}
\def\doi{8(4:19)2012}
\lmcsheading%
{\doi}
{1--28}
{}
{}
{Oct.~27, 2011}
{Nov.~29, 2012}
{}

\usepackage{enumerate}
\usepackage{hyperref}
\usepackage{epsfig}
\usepackage{fancyhdr}
\usepackage{plain}
\usepackage{mathrsfs}
\usepackage{amssymb,amsmath}
\usepackage{amsfonts}
\usepackage{amssymb}
\usepackage{stmaryrd}

\newcommand{\s}[1]{ \widetilde{#1}}
\newcommand{\lm}[1]{\displaystyle\lim_{{#1}\to\infty}}

\newcommand{\supr}[1]{\displaystyle\sup_{#1}}
\newcommand{\brck}[1]{ \llbracket{#1}\rrbracket}
\newcommand{\prth}[1]{ \llparenthesis{#1}\rrparenthesis}
\newcommand{\M}{\mathcal{M}}

\newcommand{\e}{\varepsilon}
\newcommand{\A}{\mathcal{A}}

\newcommand{\lng}{\mathcal{L}}
\newcommand{\F}{\mathcal{F}}

\newcommand{\tto}{\leftrightarrow}

\newcommand{\beq}{\begin{equation}}
\newcommand{\eeq}{\end{equation}\par\noindent}
\newcommand{\bdm}{\begin{displaymath}}
\newcommand{\edm}{\end{displaymath}}

\newcommand{\ol}{\overline}

\newcommand{\preals}{\mathbb{R}^+}
\newcommand{\rats}{\mathbb{Q}}
\newcommand{\prats}{\mathbb{Q}^+}
\newcommand{\nat}{\mathbb{N}}

\begin{document}

\title[Continuous Markovian Logics]{Continuous Markovian Logics \\Axiomatization and Quantified Metatheory\rsuper*}

\author[R.~Mardare]{Radu Mardare\rsuper a}	
\address{{\lsuper{a,c}}Department of Computer Science, Aalborg University, Selma Lagerlofs Vej 300, DK-9220 Aalborg, Denmark }	
\email{mardare@cs.aau.dk} 
\thanks{{\lsuper a}Mardare was supported by Sapere Aude: DFF-Young Researchers Grant 10-085054 of the Danish Council for Independent Research.}	

\author[L.~Cardelli]{Luca Cardelli\rsuper b}	
\address{{\lsuper b}Microsoft Research Cambridge  
7 J J Thomson Ave Cambridge CB3 0FB, UK}	
\email{luca@microsoft.com}  

\author[K.~G.~Larsen]{Kim G. Larsen\rsuper c}	
\address{\vskip-6 pt}	
\email{kgl@cs.aau.dk}  

\titlecomment{{\lsuper*}This paper is an extension of \cite{Cardelli11a} that was presented at CSL2011}


\begin{abstract}
Continuous Markovian Logic (CML) is a multimodal logic that expresses quantitative and qualitative properties of continuous-time labelled Markov processes with arbitrary (analytic) state-spaces, henceforth called continuous Markov processes (CMPs). The modalities of CML evaluate the rates of the exponentially distributed random variables that characterize the duration of the labeled transitions of a CMP. In this paper we present weak and strong complete axiomatizations for CML and prove a series of metaproperties, including the finite model property and the construction of canonical models. CML characterizes stochastic bisimilarity and it supports the definition of a quantified extension of the satisfiability relation that measures the ``compatibility'' between a model and a property. In this context, the metaproperties allows us to prove two \emph{robustness} theorems for the logic stating that one can perturb formulas and maintain ``approximate satisfaction''.
\end{abstract}

\maketitle


\section*{Introduction}\label{intro}

Many complex natural and man-made systems (e.g., biological, ecological, physical, social, financial, and computational) are modeled as stochastic processes in order to handle either a lack of knowledge or inherent randomness. These systems are frequently studied in interaction with discrete systems, such as controllers, or with interactive environments having continuous behavior. This context has motivated research aiming to develop a general theory of systems able to uniformly treat discrete, continuous and hybrid reactive systems. Two of the central questions of this research are \emph{``when do two systems behave similarly up to some quantifiable observation error?''} and \emph{``is there any (algorithmic) technique to check whether two systems have similar behaviours?''}. These questions are related to the problems of state space reduction (collapsing a model to an equivalent reduced model) and discretization (reduce a continuous or hybrid system to an equivalent discrete one), which are cornerstones in the field of stochastic systems. 

For probabilistic systems, the concept of \emph{probabilistic bisimulation} introduced by Larsen and Skou~\cite{Larsen91} is the standard concept for reasoning about probabilistic behaviours. It relates probabilistic systems with identical probabilistic behaviour, see~\cite{Panangaden09} for an expository introduction. Labelled Markov processes are the probabilistic analogs of labelled transition systems with state spaces that might be continuous. The theory of probabilistic bisimulation has been appropriately extended to cover the case of continuous-state spaces and continuous distributions~\cite{Desharnais02}. In addition, probabilistic multimodal logics (PMLs) have been used to characterize the probabilistic bisimilarity \cite{Desharnais02,Desharnais03b}. 

Despite the elegant theories supporting it, the concept of bisimulation remains too strict for applications. In modelling, the values of the parameters (rates or probabilities) are often approximated and consequently, one is interested to know whether two processes that differ by a small amount in real-valued parameters show similar (not necessarily identical) behaviours. In such cases, instead of a bisimulation relation, one needs a metric concept to estimate the degree of similarity of two systems in terms of their behaviours. The metric theory for Markov processes was initiated by Desharnais et al.~\cite{Desharnais04} and greatly developed and explored by van Breugel, Worrell and others~\cite{vanBreugel01b,vanBreugel03}. One way to do this is to relax the satisfiability relation in the logic and replace it with a function that reports the ``degree of satisfiability'' between a Markov process and a logical property. This further induces a \emph{behavioural pseudometric} on processes measuring the distance between processes in terms of their behavioural similarity. 

It was hoped that these metrics would provide a quantitative alternative to logic, but this did not happen. One reason could originate in the fact that all this ``metric reasoning'' focused exclusively on the semantics of the logic while a syntactic counterpart did not develop. When we published \cite{Cardelli11a}, which is the restricted version of this paper, there existed no attempt of understanding what a behaviour pseudometric might mean logically. We emphasized that, in the context of a completely axiomatized logic, the semantic distance between Markov processes implicitly induces, via Hausdorff metrics, a distance between probabilistic and stochastic logical properties. On this line, \cite{Cardelli11a} contains the open ideas of a research program that we have followed ever since. This research aims to understand the relation between the pseudometric space of Markov processes and the pseudometric space of logical formulas i.e., the relation between the measure of similarity of behaviours for Markov processes and the measure of provability in a corresponding stochastic/probabilistic logic. Eventually, in \cite{Larsen12b} we studied how convergence in the open ball topologies induced by the two pseudometrics ``agree to the limit'' and in \cite{Bacci12} we proposed an effective on-the-fly algorithm to compute such metric for Markov chains.   

This paper provides the corner stone for our research program: We provide weak and strong complete axiomatizations for the most general logic that express properties of stochastic labelled Markov processes. We call it \emph{continuous Markovian logic} (CML). It is similar to PML, but developed for general \emph{continuous-time and continuous-space labelled Markov processes}, henceforth continuous Markov processes (CMPs) \cite{Cardelli11a,stochspace,Cardelli11b}. CML is endowed with modal operators indexed with transition labels and positive rationals. For a label $a$ and a positive rational $r$, the formula $L^a_r\phi$ in CML expresses the fact that the rate of the $a$-transitions from the current state to the set of states satisfying $\phi$ is \emph{at least $r$}; similarly, $M^a_r\phi$ states that the rate is \emph{at most $r$}. In this respect, our logic is similar to the Aumann's system \cite{Aumann99b} developed for Harsanyi type spaces \cite{Harsanyi67}. 

In spite of their syntactic similarities, CML and PML are very different. In the probabilistic case axiomatized by Zhou in his PhD thesis \cite{Zhou07}\footnote{The semantics of \cite{Zhou07} is in terms of systems where each action is enabled with probability 1.} the two modal operators are dual, related by De Morgan laws such as $\vdash M^a_r\phi\tto L^a_{1-r}\lnot\phi$ and $\vdash L^a_r\phi\tto M^a_{1-r}\lnot\phi$, which express the fact that the probability of a transition from a state $m$ to a state satisfying $\phi$ and the probability of a transition from $m$ to a state satisfying $\lnot\phi$ sum to $1$. In the stochastic case the two modalities are independent. Moreover, there exists no sound equivalence of type $\lnot X^a_r\phi\tto Y^a_s\lnot\phi$ for $X,Y\in\{L,M\}$, since the rate of the transitions from $m$ to the set of states satisfying $\phi$ is not related to the rate of the transitions from $m$ to the set of states satisfying $\lnot\phi$. Hence, for CML formulas, no positive normal forms can be defined. Obviously, these differences are also reflected in the complete axiomatizations that we present both for CML and for its fragment without $M^a_r$-operators. Many axioms of PML, such as $\vdash L^a_r\top$ or $\vdash L_r^a\phi\to\lnot L^a_s\lnot\phi$ for $r+s<1$, are not sound for CMPs. 

Another important difference between PML and CML regards their metaproperties. The proof of the finite model property for PML, done in \cite{Zhou07}, on which many other results such as decidability or weak completeness are centered, relays on the fact that in $[0,1]$ there are finitely many rationals of a given denominator. When one moves from probabilistic to stochastic processes, i.e., from probabilistic distributions to arbitrary distributions, this proof cannot be reproduced and for this reason decidability and weak completeness for CML remained open problems for some time. In this paper we prove the finite model property for CML using a nontrivial variation of Zhou's construction. This is one of the major contribution of this paper. 

The construction of a finite model for a consistent CML-formula is also relevant in the context of pseudometrics. It provides an approximation techniques to evaluate the quantitative extension of satisfiability relation induced by the behavioural pseudometric. Formally, the \emph{quantitative satisfiability} is a function $d:\mathfrak P(\A)\times \lng(\A)\to\preals$ that associates to a CMP $P\in\mathfrak P(\A)$ and a CML formula $\phi\in\lng(\A)$ a value $d(P,\phi)\in\preals$ that measures the distance between $P$ and the set of processes satisfying $\phi$: $P\models\phi$ iff $d(P,\phi)=0$. This function induces a distance on the space of logical formulas, $\ol d:\lng(\A)\times\lng(\A)\to\preals$ defined by
$$\ol d(\phi,\phi')=\supr{P\in\mathfrak P(\A)}|d(P,\phi)-d(P,\phi')|.$$
We observe that, in the context of a complete axiomatization, the distance $\ol d(\phi,\phi')$ measures the similarity between logical formulas in terms of provability: $\phi$ and $\phi'$ are at distance 0 in $\ol d$ if they can both be entailed from the same consistent theories. In this context we prove the \emph{Strong Robustness Theorem}: $$d(P,\phi)\leq d(P,\phi')+\ol d(\phi,\phi').$$ 
In case that $\ol d(\phi,\phi')$ is not computable or it is very expensive, one can use our finite model construction to approximate its value. Let $\s{d}(\phi,\phi')=max\{|d(P,\phi)-d(P,\phi')|, P\in\Omega_p[\psi]\}$, where $\Omega_p[\psi]$ is the finite model (it contains a finite set of processes) constructed for a consistent formula $\psi$ for which both $\phi$ and $\phi'$ belong to the Fischer-Ladner closure of $\psi$ and $p\in\nat$ is a special parameter involved in the construction of the finite model. This guarantees the \emph{Weak Robustness Theorem}: $$d(P,\phi)\leq d(P,\phi')+\s{d}(\phi,\phi')+2/p.$$ Using this second theorem, one can evaluate $d(P,\phi)$ from the value of $d(P,\phi')$. Of course, the accuracy of this approximation depends on the similarity between $\phi$ and $\phi'$ from a provability perspective, which influences both the distance $\s{d}(\phi,\phi')$ and the parameter $p$ of the finite model construction. 
\bigskip

To summarize, the achievements of this paper are as follows.
\begin{iteMize}{$\bullet$}
\item We introduce Continuous Markovian Logic, a modal logic that expresses quantitative and qualitative properties of continuous Markov processes. CML is endowed with operators that approximate the labelled transition rates of CMPs and allows us to reason on approximated properties. This logic characterizes the stochastic bisimulation of CMPs.
\item We present weak and strong completeness results for CML and for its fragment without $M^a_r$ operators. These are very different from the similar probabilistic cases, due to the structural differences between probabilistic and stochastic models; and the differences are reflected by the axioms.
\item We prove the finite model properties for CML and its restricted fragment. The construction of a finite model for a consistent formula is novel in the way it exploits the Archimedian properties of positive rationals. We extend the construction to define canonic models for the two logics.
\item We define a distance between logical formulas related to the distance between a model and a formula proposed in the literature for probabilistic systems. The organization of the space of logical formulas as a pseudometric space, with a topology sensitive to provability, is a novelty in the field of metric semantics. This structure guarantees the strong robustness theorem.
\item We show that the complete axiomatization and the finite model construction can be used to approximate the syntactic distance between formulas. This idea opens new research perspectives on the direction of designing algorithms to estimate such distances within given errors.   
\end{iteMize}

\noindent\textbf{The structure of the paper.} We end the introduction with a section that comprises some preliminary concepts and notations used in the paper. Section \ref{CMP} introduces CMPs and their bisimulation. In Section \ref{logic} we define the logic CML and its semantics. Section \ref{axioms} is dedicated to the weak completeness of the fragment of CML without $M^a_r$ operators and to the proof of the finite model property for this fragment. In Section \ref{axioms1} we extend the weak completeness proof and the construction of the finite model for consistent formulas to the entire logic. Section \ref{canonic} extends the axiomatizations to prove the strong completeness and the existence of canonical models for the two logics. Section \ref{characterize} introduces the metric semantics and the results related to metrics and bisimulation; in this section we present and discuss the robustness theorems. The paper also contains a conclusive section where we comment on the new research directions opened by this paper.

\section*{Preliminary definitions and notations}\label{sec1}

In this section we establish the terminology used in the paper. Most of the notation is standard. We also present some classic results that play an important role in the economy of the paper. However, we assume that the reader is familiar with the basic terminology of set theory, topology and measure theory.


\subsection*{Sets, Relations, Functions}

For arbitrary sets $A$ and $B$, $2^A$ denotes the \emph{powerset} of $A$, $A\uplus B$ their \emph{disjoint union} and $[A\to B]$ the class of functions from $A$ to $B$. 

If $f:A\to B$, we denote by $f^{-1}:2^B\to 2^A$ the \emph{inverse mapping} of $f$ defined, for arbitrary $B'\subseteq B$ by $f^{-1}(B')=\{a\in A\mid f(a)\in B'\}$. If $0\in B$, the \emph{kernel of $f$} is the set $ker(f)=f^{-1}(\{0\})$ 

Given a set $A$ and a relation $\mathfrak R\subseteq A\times A$, the set $A'\subseteq A$ is \emph{$\mathfrak R$-closed} iff $$\{a\in A~|~\exists a'\in A', (a',a)\in\mathfrak R\}\subseteq A'.$$
If $\Sigma\subseteq 2^A$ is a set of subsets of $A$, then $\Sigma(\mathfrak R)$ denotes the set of $\mathfrak R$-closed elements of $\Sigma$.


\subsection*{Measurable Spaces}
In what follows we introduce a few concepts from measure theory and state a few results that are essential for our paper. For their proofs and more related results the reader is referred to \cite{Billingsley95}.

\begin{defi}
Let $M$ be an arbitrary set.

\begin{iteMize}{$\bullet$}
\item A nonempty family of subsets $\Pi\subseteq 2^M$ closed under finite intersection is called a \emph{$\pi$-system}. 
\item A nonempty family of subsets $\F\subseteq 2^M$ that contains $M$ and is closed under complement and finite intersection is called a \emph{field}. 
\item A nonempty family of subsets $\mathcal S\subseteq 2^M$ that contains $\emptyset$ and it is closed under finite intersection is called a \emph{semiring}, if for arbitrary $A,B\in\mathcal S$ such that $A\subseteq B$, $B\setminus A$ is a finite union of elements in $\mathcal S$. 
\item A nonempty family of subsets $\Sigma\subseteq 2^M$ is a \emph{$\sigma$-algebra} over $M$ if it contains $M$ and is closed under complement and countable union. 
\end{iteMize}
If $\Sigma$ is a $\sigma$-algebra over $M$, the tuple $(M,\Sigma)$ is called a \emph{measurable space}, the elements of $\Sigma$ \emph{measurable sets} and $M$ is called the \emph{support-set} of the space.
\end{defi}

If $\Omega\subset 2^M$ is a nonempty family of subsets of $M$, the \emph{$\sigma$-algebra generated by $\Omega$}, denoted $\Omega^\sigma$ is the smallest $\sigma$-algebra containing $\Omega$.

\begin{defi}
A \emph{measure} on a measurable space $\M=(M,\Sigma)$ is a countably additive function $\mu:\Sigma\to\mathbb R^+$ such that $\mu(\emptyset)=0$. We use $\Delta(M,\Sigma)$ to denote the set of measures on $(M,\Sigma)$. 
\end{defi}

Given a measurable space $\M=(M,\Sigma)$, we organize the class $\Delta(M,\Sigma)$ of measures as a measurable space by considering the $\sigma$-algebra $\mathfrak F$ generated, for arbitrary $S\in\Sigma$ and $r>0$, by the sets $F(S,r)=\{\mu\in\Delta(M,\Sigma):\mu(S)\geq r\}.$

\begin{defi}
Given two measurable spaces $(M,\Sigma)$ and $(N,\Omega)$, a mapping $f:M\to N$ is \emph{measurable} if $\mbox{for any }T\in\Omega, f^{-1}(T)\in\Sigma.$ We use $\llbracket M\to N\rrbracket$ to denote the class of measurable mappings from $(M,\Sigma)$ to $(N,\Omega)$.
\end{defi}

Now we state a few results that are fundamental in the construction of the finite and the canonic model for CML. 

\begin{thm}[\cite{Billingsley95}, Theorem 10.3]\label{A} 
Suppose that $\Pi\subseteq 2^M$ is a $\pi$-system with $M\in\Pi$ and $\mu,\nu$ are two measures on $(M,\Pi^\sigma)$. If $\mu$ and $\nu$ agree on all the sets in $\Pi$, then they agree on $\Pi^\sigma$.
\end{thm}

\begin{thm}[\cite{Billingsley95}, Theorem 11.3]\label{A1}
If $\mu$ is a set function on a semiring $\mathcal S$ with values in $[0,\infty]$ such that $\mu(\emptyset)=0$, $\mu$ is finitely additive and countably subadditive, then $\mu$ extends to a measure on $\mathcal S^\sigma$.
\end{thm}

\begin{defi}
A set function $\mu$ defined on a field $\mathcal F$ is \emph{continuous from above in 0} if for any decreasing sequence $(A_i)_{i\in\mathbb N}\subseteq\mathcal F$ such that $\displaystyle\bigcap_{i\in I}A_i=\emptyset$, $\lm{i}\mu(A_i)=0$.
\end{defi}

\begin{thm}[\cite{Billingsley95}, Example 2.10]\label{A2}
If $\mu$ is a set function on a field $\mathcal F$ such that $\mu(\emptyset)=0$, $\mu$ is finitely additive and it is continuous from above in 0, then $\mu$ is countably additive on $\mathcal F$.
\end{thm}

These previous results allow us to prove the next lemma that is the key result in our model constructions.

\begin{lem}\label{semiring}
If $\mathcal F$ is a field, then any set function $\mu:\mathcal F\to\mathbb R^+$ that is finitely additive, continuous from above in 0 and such that $\mu(\emptyset)=0$ can be uniquely extended to a measure on $\mathcal F^\sigma$.
\end{lem}

\proof
Because $\mathcal F$ is a field, Theorem \ref{A2} guarantees that $\mu$ is countably additive on $\mathcal F$, hence countably subadditive. It is trivial to verify that any field is a semiring. Now we apply Theorem \ref{A1}, since $\mathcal F$ is also a semiring, and we obtain that $\mu$ extends to a measure on $\mathcal F^\sigma$. The uniqueness derives from Theorem \ref{A}.
\qed


\subsection{Analytic Spaces}
The analytic spaces play a central role in this paper. They are restrictions of general measure spaces but, however they form a very wide class. The analytic spaces have some remarkable properties that are needed for some of our results. For instance, the proof of the logical characterization of bisimulation cannot be done for arbitrary spaces \cite{Desharnais02}, and a good source for the unprovability of logical characterization of bisimulation for arbitrary spaces can be found in \cite{Terraf11}. In what follows we only present the definition of Polish and analytic spaces. For a complete exposition on this topic, the reader is referred to~\cite{Dudley89} or~\cite{Arveson76}.

\begin{defi}
A \textbf{Polish} space is the topological space underlying a complete, separable metric space; i.e.\  it has a countable dense subset or
equivalently a countable base of open sets.
\end{defi}

\begin{defi}
An \textbf{analytic} space is the image of a Polish space under a continuous function between Polish spaces.  
\end{defi}


\section{Continuous Markov processes}\label{CMP}

In this section we introduce the continuous-time Markov processes \cite{Cardelli11a}, henceforth, continuous Markov processes (CMPs), which are models of stochastic systems with arbitrary (possible continuous) state space and continuous-time transitions. Such systems were introduced for the first time in \cite{Desharnais03b}. In this paper we use a different definition proposed in \cite{Cardelli11a,Cardelli11b}, which exploits an equivalence between the
definitions of Harsanyi type spaces \cite{Harsanyi67,Moss04} and a coalgebraic view of labelled Markov processes~\cite{deVink99} described, for example, by Doberkat~\cite{Doberkat07}.  

The CMPs are defined for a fixed countable set $\A$ of \emph{transition labels} representing the types of interactions with the environment. If $a\in\A$, $m$ is the current state of the system and $N$ is a measurable set of states, the function $\theta(a)(m)$ is a measure on the state space and $\theta(a)(m)(N)\in\mathbb R^+$ represents the \emph{rate} of an exponentially distributed random variable that characterizes the duration of an $a$-transition from $m$ to arbitrary $n\in N$. Indeterminacy in such systems is resolved by races between events executing at different rates. 

\begin{defi}[Continuous Markov processes]
Given an analytic space $(M,\Sigma)$, where $\Sigma$ is the Borel algebra generated by the topology, an \emph{$\A$-continuous Markov kernel} is a tuple $\M=(M,\Sigma,\theta)$, where 
$\theta:\A\to\llbracket M\to\Delta(M,\Sigma)\rrbracket.$  $M$ is the support set of $\M$ denoted by $supp(\M)$. If $m\in M$, $(\M,m)$ is an \emph{$\A$-continuous Markov process}.
\end{defi}

Notice that $\theta(a)$ is a measurable mapping between $(M,\Sigma)$ and $(\Delta(M,\Sigma),\mathfrak F)$, where $\mathfrak F$ is the $\sigma$-algebra on $\Delta(M,\Sigma)$ defined in the preliminaries. This condition is equivalent to the conditions on the two-variable \emph{rate function} used in \cite{Panangaden09} to define continuous Markov processes. For the proof of this equivalence, see  e.g. Proposition 2.9, of \cite{Doberkat07}.

\bigskip

In the rest of the paper we assume that the set of transition labels $\A$ is fixed. We denote by $\mathfrak M(\A)$ the class of $\A$-continuous Markov kernels (CMKs) and we use $\M,\M_i,\M'$ to range over $\mathfrak M(\A)$. We denote by $\mathfrak P(\A)$ the set of $\A$-continuous Markov processes (CMPs) and we use $P,P_i,P'$ to range over $\mathfrak P(\A)$.
 
The stochastic bisimulation for CMPs follows the line of Larsen and Skou's probabilistic bisimulation \cite{Larsen91}, adapted to continuous state-spaces by Desharnais et al.~\cite{Desharnais02,Desharnais03}. Recall that in the next definition $\Sigma(\mathfrak R)$ denotes the set of $\mathfrak R$-closed measurable sets (see preliminaries).

\begin{defi}[Stochastic Bisimulation]
Given $\M=(M,\Sigma,\theta)\in\mathfrak M(\A)$, a \emph{rate-bi\-simu\-lation} on $\M$ is an equivalence relation $\mathfrak R\subseteq M\times M$ such that $(m,n)\in\mathfrak R$ implies that for any $C\in\Sigma(\mathfrak R)$ and any $a\in\A$,
$$\theta(a)(m)(C)=\theta(a)(n)(C).$$
Two processes $(\M,m)$ and $(\M,n)$ are \emph{stochastic bisimilar}, written $m\sim_\M n$, if they are related by a rate-bisimulation relation.
\end{defi}

Observe that, for any $\M\in\mathfrak M(\A)$ there exist rate-bisimulation relations as, for instance, is the identity relation on $\M$; the \emph{stochastic bisimilarity} is the largest rate-bisimulation. 
 
\begin{defi}[Disjoint union of Markov kernels]
The disjoint union of $\M=(M,\Sigma,\theta)$ and
$\M'=(M',\Sigma',\theta')$ in $\mathfrak
M(\A)$ is given by $\M''=(M'',\Sigma'',\theta'')$ such
that $M''=M\uplus M'$, $\Sigma''$ is generated by
$\Sigma\uplus\Sigma'$ and for any $a\in\mathcal A$, $N\in\Sigma$ and
$N'\in\Sigma'$,
$$\begin{array}{ll}
\theta''(a)(m)(N\uplus N')= & \left\{
\begin{array}{ll}
\theta(a)(m)(N) & \textrm{if } m\in M\\
\theta'(a)(m)(N') & \textrm{if }m\in M'\\
\end{array}\right. \\
\end{array}$$
The disjoint union of $\M$ and $\M'$ is denoted by $\M\uplus\M'$.
\end{defi}

Notice that $\M\uplus\M'\in\mathfrak M(\A)$. If $m\in M$ and $m'\in M'$, we say that $(\M,m)$ and $(\M',m')$ are \emph{bisimilar }written $(\M,m)\sim (\M',m')$ whenever $m\sim_{\M\uplus\M'}m'$.


\section{Continuous Markovian Logics}\label{logic}

In this section we introduce a class of modal logics, henceforth, the continuous Markovian logics (CMLs) with semantics defined in terms of CMPs. In addition to the Boolean operators, these logics are endowed with \emph{stochastic modal operators} that evaluate, from below and from above, the rates of the transitions of a CMP. 

For $a\in\A$ and $r\in\mathbb Q_+$, $L^a_r\phi$ is satisfied by $(\M,m)\in\mathfrak P(\A)$ whenever the rate of the $a$-transition from $m$ to the class of the states satisfying $\phi$ is \emph{at least $r$}; symmetrically, $M^a_r\phi$ is satisfied when this rate is \emph{at most $r$}. CMLs extends the probabilistic logics \cite{Aumann99b,Larsen91,Heifetz01,Zhou07,Fagin94} to stochastic domains. The structural similarities between the probabilistic and the stochastic models are not preserved when we consider the logic. Because we focus on general measures instead of probabilistic measures in the definition of the transition systems, many of the axioms of probabilistic logics are not sound for stochastic semantics. This is the case, for instance, with $\vdash L^a_r\top$ or $\vdash L_r^a\phi\to\lnot L^a_s\lnot\phi$ for $r+s<1$ which are proposed in \cite{Heifetz01,Zhou07}. Moreover, while in probabilistic settings the operators $L^a_r$ and $M^a_s$ are dual, satisfying the De Morgan laws $\vdash M^a_r\phi\tto L^a_{1-r}\lnot\phi$ and $\vdash L^a_r\phi\tto M^a_{1-r}\lnot\phi$, they became independent in stochastic semantics. For this reason, in the next sections we study two CML logics with complete axiomatizations, $\lng(\A)$ involving only the stochastic operators of type $L^a_r$ and $\lng(\A)^+$ that contains both $L^a_r$ and $M^a_s$.

In what follows we use the same set $\A$ of labels used with CMPs. 
\begin{defi}[Syntax]
Given a countable set $\A$, the formulas of $\lng(\A)$ and $\lng^+(\A)$ respectively are introduced by the following grammars, for arbitrary $a\in\mathcal A$ and $r\in\mathbb Q_+$.
$$\lng(\A):~~~\phi:=\top~|~\lnot\phi~|~\phi\land\phi~|~L^a_r\phi,$$ $$\lng^+(\A):~~~\phi:=\top~|~\lnot\phi~|~\phi\land\phi~|~L^a_r\phi~|~M^a_r\phi.$$
\end{defi}

In addition, we assume all the Boolean operators, including $\bot=\lnot\top$, as well as the derived operator $E^a_r\phi=L^a_r\phi\land M^a_r\phi$.

The semantics of $\lng(\A)$ and $\lng^+(\A)$, called in this paper \emph{Markovian semantics}, are defined by the \emph{satisfiability relation} for arbitrary $(\M,m)\in\mathfrak P(\A)$ with $\M=(M,\Sigma,\theta)\in\mathfrak M(\A)$, by:

$\M,m\models \top$ always,

$\M,m\models\lnot\phi$ iff it is not the case that $\M,m\models\phi$,

$\M,m\models\phi\land\psi$ iff $\M,m\models\phi$ and $\M,m\models\psi$,

$\M,m\models L^a_r\phi$ iff $\theta(a)(m)(\llbracket \phi\rrbracket_\M)\geq r$,

$\M,m\models M^a_r\phi$ iff $\theta(a)(m)(\llbracket \phi\rrbracket_\M)\leq r$,
\\where $\llbracket \phi\rrbracket_\M=\{m\in M|\M,m\models\phi\}$.

When it is not the case that $\M,m\models\phi$, we write $\M,m\not\models\phi$. 

From here we get the obvious rules for the derived operators:

$\M,m\not\models\bot$ always,

$\M,m\models E^a_r\phi$ iff $\theta(a)(m)(\llbracket\phi\rrbracket_\M)=r$. 

Notice that $E^a_r\phi$ characterizes the process that can do an $a$-transition with exactly the rate $r$ to the set of processes characterized by $\phi$. In the stochastic case $L^a_r$, $M^a_r$ and $E^a_r$ are mutually independent. We chose not to study a Markovian logic that involves only the $E^a_r$ operators because in many applications we do not know the exact rates of the transitions and it is more useful to work with approximations such as $M^a_r$ or $L^a_r$.

Observe that the semantics of $L^a_r\phi$ and $M^a_r\phi$ are well defined only if $\llbracket \phi\rrbracket_\M$ is measurable. This is guaranteed by the fact that $\theta(a)$ is a measurable mapping between $(M,\Sigma)$ and $(\Delta(M,\Sigma),\mathfrak F)$, as shown in the next lemma.

\begin{lem}\label{measurability}
For any $\phi\in\lng^+(\A)$ and any $\M=(M,\Sigma,\theta)\in\mathfrak M(\A)$, $\llbracket\phi\rrbracket_\M\in\Sigma$.
\end{lem}

\proof 
Induction on $\phi$: the Boolean cases are trivial since $\brck{\top}_\M=M$ is measurable, and the measurability is closed with respect to finite intersections and complement.

\textbf{The case $\phi= L^a_r\psi$:} the inductive hypothesis guarantees that $\llbracket\psi\rrbracket_\M\in\Sigma$, hence, $\{\mu\in\Delta(M,\Sigma)|\mu(\llbracket\psi\rrbracket_\M)\geq r\}$ is measurable in $\Delta(M,\Sigma)$. Because $\theta(a)$ is a measurable mapping, we obtain that $\llbracket L^a_r\psi\rrbracket_\M=(\theta(a))^{-1}(\{\mu\in\Delta(M,\Sigma)|\mu(\llbracket\psi\rrbracket_\M)\geq r\})$ is measurable. 

Similarly it can be proved for $\phi=M^a_r\psi$.
\qed 

\begin{cor}
For any $\phi\in\lng(\A)$ and any $\M=(M,\Sigma,\theta)\in\mathfrak M(\A)$, $\llbracket\phi\rrbracket_\M\in\Sigma$.
\end{cor}

As usually, we say that a formula $\phi$ is \emph{satisfiable} if there exists $\M=(M,\Sigma,\theta)\in\mathfrak M(\A)$ and $m\in M$ such that $\M,m\models\phi$. $\phi$ is \emph{valid}, denoted by $\models\phi$, if $\lnot\phi$ is not satisfiable.


\section{Weak Completeness for $\lng(\A)$}\label{axioms}

In this section we present a Hilbert-style axiomatization for $\lng(\A)$ and we prove its soundness and weak completeness for the Markovian semantics. The axioms and rules, collected in Table \ref{AS} are given for propositional variables $\phi,\psi\in\lng(\A)$, for arbitrary $a\in\mathcal A$ and $s,r\in\mathbb Q^+$. In addition we also assume the axiomatization of the classic propositional logic.

\begin{table}[!h]
$$
   \begin{array}{ll}
        \mbox{(A1):} & \vdash L^a_0\phi\\
        \mbox{(A2):} & \vdash L^a_{r+s}\phi\to L^a_r\phi\\
        \mbox{(A3):} & \vdash L^a_r(\phi\land\psi)\land L^a_s(\phi\land\lnot\psi)\to L^a_{r+s}\phi\\
        \mbox{(A4):} & \vdash \lnot L^a_r(\phi\land\psi)\land \lnot L^a_s(\phi\land\lnot\psi)\to \lnot L^a_{r+s}\phi\\
        \mbox{(R1):} & \mbox{If }\vdash\phi\rightarrow\psi\mbox{ then }\vdash L^a_r\phi\to L^a_r\psi\\
        \mbox{(R2):} & \{L^a_r\phi\mid r<s\}\vdash L^a_s\phi\\
        \mbox{(R3):} & \{L^a_r\phi\mid r>s\}\vdash\bot\\
   \end{array}
$$\caption{\label{AS}The axiomatic system of $\lng(\A)$}
\end{table}

Axiom (A1) guarantees that the rate of any transition is at least $0$ and encodes the fact that the measure of any set cannot be negative. (A2) states that if a rate is at least $r+s$ then it is at least $r$. (A3) and (A4) encode the additive properties of measures for disjoint sets: $\brck{\phi\land\psi}$ and $\brck{\phi\land\lnot\psi}$ are disjoint sets of processes such that $\brck{\phi\land\psi}\cup\brck{\phi\land\lnot\psi}=\brck\phi$. The rule (R1) establishes the monotonicity of $L^a_r$. In this axiomatic system we have two infinitary rules, (R2) and (R3). (R2) reflects the Archimedian property of rationals: if the rate of a transition from a state to a given set of states is at least $r$ for any $r<s$, then it is at least $s$. (R3) eliminates the possibility of having transitions at infinite rates.

In Table \ref{ASPL} below we present the similar axiomatization that Zhou proposes in \cite{Zhou07} for probabilistic logic and Harsanyi type spaces. Notice the main differences between the two systems: the axioms (B2), (B3) and (B4) of probabilistic logic are not sound for the Markovian semantics and this changes the entire structure of the provability relation. There are also important differences between the axioms (A3) and (A4) on one hand and (B4), (B5) on the other hand. In the probabilistic case there exist De Morgan relations between the two modal operators stated by (B3). We will see in the next section that a similar relation is impossible in the stochastic case. 

\begin{table}[!h]
$$
   \begin{array}{ll}
        \mbox{(B1):} & \vdash L^a_0\phi\\
        \mbox{(B2):} & \vdash L^a_r\top\\
        \mbox{(B3):} & \vdash L^a_r\phi\tto M^a_{1-r}\lnot\phi\\
        \mbox{(B4):} & \vdash L^a_r\phi\to \lnot L^a_s\lnot\phi,~~~r+s>1\\
        \mbox{(B5):} & \vdash L^a_r(\phi\land\psi)\land L^a_s(\phi\land\lnot\psi)\to L^a_{r+s}\phi,~~~r+s\leq 1\\
        \mbox{(B6):} & \vdash \lnot L^a_r(\phi\land\psi)\land \lnot L^a_s(\phi\land\lnot\psi)\to \lnot L^a_{r+s}\phi,~~~r+s\leq 1\\
        \mbox{(S1):} & \mbox{If }\vdash\phi\rightarrow\psi\mbox{ then }\vdash L^a_r\phi\to L^a_r\psi\\
        \mbox{(S2):} & \{\lnot M^a_r\phi\mid r<s\}\vdash L^a_s\phi
    \end{array}
$$\caption{\label{ASPL}The axiomatic system of PML}
\end{table}

As usual, we say that a formula $\phi$ is \emph{provable}, denoted by $\vdash\phi$, if it can be proved from the given axioms and rules. We say that $\phi$ is \emph{consistent}, if $\phi\to\bot$ is not provable. Given a set $\Phi$ of formulas, we say that $\Phi$ proves $\phi$, $\Phi\vdash\phi$, if from the formulas of $\Phi$ and the axioms one can prove $\phi$. $\Phi$ is \emph{consistent} if it is not the case that $\Phi\vdash\bot$; $\Phi$ is \emph{finite-consistent} if any finite subset of it is consistent. For a sublanguage $\lng\subseteq\lng(\mathcal A)$, we say that $\Phi$ is $\lng$-maximal if no formula from $\lng$ can be added to $\Phi$  without making it inconsistent; $\Phi$ is \emph{$\lng$-maximally consistent} if $\Phi$ is consistent and $\lng$-maximal.

\begin{thm}[Soundness]
The axiomatic system of $\lng(\A)$ is sound for the Markovian semantics, i.e., for any $\phi\in\lng(\A)$, if $\vdash\phi$ then $\models\phi$.
\end{thm}

\proof
As usual, the soundness proof consists in proving that each axiom is sound and that the rules preserve soundness. This is sufficient to guarantee that we can only derive sound consequences from sound hypothesis. In what follows we consider an arbitrary $(\M,m)\in\mathfrak P(\A)$ with $\M=(M,\Sigma,\theta)$.

\textbf{(A1):} We have $\models L^a_0\phi$ since $\theta(a)(m)(\brck{\phi})\geq 0$ for any $\phi$.

\textbf{(A2):} We have $\models L^a_{r+s}\phi\to L^a_r\phi$ since $\M,m\models L^a_{r+s}\phi$ is equivalent to $\theta(a)(m)(\brck{\phi})\geq r+s$ implying $\theta(a)(m)(\brck{\phi})\geq r$, i.e., $\M,m\models L^a_r\phi$.

\textbf{(A3):} Suppose that $\M,m\models L^a_r(\phi\land\psi)\land L^a_s(\phi\land\lnot\psi)$. Then, $\theta(a)(m)(\brck{\phi\land\psi})\geq r$ and $\theta(a)(m)(\brck{\phi\land\lnot\psi})\geq s$. But since $\brck{\phi\land\psi}$ and $\brck{\phi\land\lnot\psi}$ are disjoint sets of processes such that $\brck{\phi\land\psi}\cup\brck{\phi\land\lnot\psi}=\brck\phi$, $\theta(a)(m)(\brck\phi)=\theta(a)(m)(\brck{\phi\land\psi})+\theta(a)(m)(\brck{\phi\land\lnot\psi})$. Hence, $\theta(a)(M)(\brck\phi)\geq r+s$, i.e., $\M,m\models L^a_{r+s}\phi$.

\textbf{(A4):} Suppose that $\M,m\models \lnot L^a_r(\phi\land\psi)\land \lnot L^a_s(\phi\land\lnot\psi)$. Then, $\theta(a)(m)(\brck{\phi\land\psi})< r$ and $\theta(a)(m)(\brck{\phi\land\lnot\psi})< s$. But since $\brck{\phi\land\psi}$ and $\brck{\phi\land\lnot\psi}$ are disjoint sets of processes such that $\brck{\phi\land\psi}\cup\brck{\phi\land\lnot\psi}=\brck\phi$, $\theta(a)(m)(\brck\phi)=\theta(a)(m)(\brck{\phi\land\psi})+\theta(a)(m)(\brck{\phi\land\lnot\psi})$. Hence, $\theta(a)(M)(\brck\phi)< r+s$, i.e., $\M,m\models \lnot L^a_{r+s}\phi$.

\textbf{(R1):} If $\models\phi\to\psi$, then $\brck\phi\subseteq\brck\psi$. Suppose that $\M,m\models L^a_r\phi$. Then, $\theta(a)(m)(\brck\phi)\geq r$. Since $\brck\phi\subseteq\brck\psi$, we derive that $\theta(a)(m)(\brck\phi)\leq \theta(a)(m)(\brck\psi)$, hence, $\theta(a)(m)(\brck\psi)\geq r$ implying $\M,m\models L^a_s\psi$.

\textbf{(R2):} Suppose that for all $r<s$, $\M,m\models L^a_r\phi$, i.e., for all $r<s$, $\theta(a)(m)(\brck\phi)\geq r$. Using the Archimedean property of rationals, we derive that $\theta(a)(m)(\brck\phi)\geq s$. Hence, $\M,m\models L^a_s\phi$.

\textbf{(R3):} Suppose that for all $r>s$, $\M,m\models L^a_r\phi$, i.e., for all $r>s$, $\theta(a)(m)(\brck\phi)\geq r$. But then, $\theta(a)(m)(\brck\phi)=\infty$ - impossible since $\brck\phi$ is measurable and the measure is always finite. Hence, there exists no process$(\M,m)$ with this property, i.e., $\{L^a_r\phi\mid r>s\}$ is inconsistent.
\qed


In the rest of this section we prove the weak completeness of the axiomatic system for the Markovian semantics, i.e., that any valid formula can be proved from the given axioms and rules. In order to do this, we prove the finite model property for our logic stating that any consistent $\lng(\A)$-formula has a finite model. 

To prove this, we will construct a model $(\M_\psi,\Gamma)\in\mathfrak P(\A)$ for an arbitrary consistent formula $\psi\in\lng(\A)$, where $supp(\M_\psi)$ is a finite set of $\lng(\A)$-consistent sets of formulas. As usual with the filtration method, the key result is the Truth Lemma stating that $\psi\in \Gamma$ iff $\M_\psi,\Gamma\models\psi$. A similar construction has been proposed in \cite{Zhou07} for probabilistic logic. However, for the probabilistic case the proof relies on the fact that there exists a finite set of rationals of a fixed denominator within $[0,1]$. Since in the stochastic case the rates range on $[0,\infty)$, this property cannot be used and instead we will have to handle a more complicated construction.

\subsection*{Notations.} 

Before proceeding with the construction, we fix some notations. 

For $n\in\mathbb N$, $n\neq 0$, let $\mathbb Q_n=\{\frac{p}{n} : p\in\mathbb N\}$. If $S\subseteq\mathbb Q$ is finite, the \emph{granularity of $S$}, $gr(S)$, is the least common denominator of the elements of $S$.

Consider an arbitrary formula $\phi\in\lng(\A)$ and let $R\subseteq\prats$ be the set of all $r\in\prats$ such that $r$ is the index of an operator $L^a_r$ present in the syntax of $\phi$.
\begin{iteMize}{$\bullet$}
\item The \emph{granularity of} $\phi\in\lng$, denoted by $gr(\phi)$ is defined by $gr(\phi)=gr(R)$
\item The \emph{upper bound of $\phi$}, denoted by $max(\phi)$ is defined by $max(\phi)=max(R)$.
\item The \emph{modal depth of} $\phi$, denoted by $md(\phi)$, is defined inductively by 
$$\begin{array}{ll}
md(\phi)= & \left\{
\begin{array}{ll}
0, & \textrm{ if } \phi=\top\\
md(\psi), & \textrm{ if } \phi=\lnot\psi\\
max\{md(\psi),md(\psi')\}, & \textrm{ if }\phi=\psi\land\psi'\\
md(\psi)+1, &\textrm{ if } \phi=L^a_r\psi
\end{array}\right. \\
\end{array}$$
\item The \emph{actions of} $\phi$ is the set $act(\phi)\subseteq\A$ of indexes $a\in\A$ of the operators $L^a_r$ present in the syntax of $\phi$.\smallskip
\end{iteMize}

\noindent For arbitrary $n\in\mathbb N$ and $A\subseteq\A$, let $\lng_n(A)$ be the sublanguage of $\lng(\A)$ that uses only modal operators $L^a_r$ with $r\in\mathbb Q_n$ and $a\in A$. 

For $\Lambda\subseteq\lng(\A)$, let $[\Lambda]_n=\Lambda\cup\{\phi\in\lng_n(\A):\Lambda\vdash\phi\}$.

The next lemma has a central role in the finite model construction.

\begin{lem}\label{construction1}
If $\Lambda\subseteq\lng(\A)$ is a finite consistent set of formulas, then for any $\phi\in\lng(\A)$ and $a\in\A$,
\begin{enumerate}[\em(1)]
\item there exists $r\in\prats$ such that $\Lambda\cup\{\lnot L^a_r\phi\}$ is consistent;
\item there exists $sup\{r\in\prats\mid\Lambda\vdash L^a_r\phi\}$.
\end{enumerate}
\end{lem}

\proof\hfill
\begin{enumerate}[(1)]
\item Suppose that there exists no $r\in\prats$ such that $\Lambda\cup\{\lnot L^a_r\phi\}$ is consistent. Then, $\Lambda\vdash \{L^a_r\phi\mid\mbox{ for any }r>0\}$. Using (R3), this implies the inconsistency of $\Lambda$ - contradiction.

\item Suppose that $sup\{r\in\prats\mid\Lambda\vdash L^a_r\phi\}$ does not exists. Then, $\Lambda\vdash \{L^a_r\phi\mid\mbox{ for any }r>0\}$ and, again, we derive from (R3) the inconsistence of $\Lambda$ - contradiction.
\qed
\end{enumerate}

\subsection*{The construction of a finite model for a consistent $\lng(\A)$-formula}\hfill

We have now the necessary prerequisite to describe our construction. Consider a consistent formula $\psi\in\lng(\A)$ with $gr(\psi)=n$ and $act(\psi)=A$. 
Let $$\lng[\psi]=\{\phi\in\lng_n(A)\mid max(\phi)\leq max(\psi), md(\phi)\leq md(\psi)\}.$$

In this construction, $\lng[\psi]$ plays a similar role to the Fischer-Ladner closure in the filtration method: we construct a CMK $\M_\psi\in\mathfrak M(\A)$ with $supp(\M_\psi)$ being the set of $\lng[\psi]$-maximally consistent sets of formulas\footnote{In fact, for the economy of the construction, we need to work with slightly larger sets that include the $\lng[\psi]$-maximally consistent sets.}. And for this model, which is finite since $\lng[\psi]$ is finite modulo propositional equivalence, we will prove the Truth Lemma: $$\mbox{for any }\phi\in\lng[\psi], \phi\in\Gamma\mbox{ iff }\M_\psi,\Gamma\models\phi.$$

To compute our task, we need to define, for each $\phi\in\lng[\psi]$, each $a\in A$ and each $\Gamma\in supp(\M_\psi)$, the value of $\theta(a)(\Gamma)(\brck\phi)$ and eventually to prove that this function satisfies the requirements of a transition function of a CMP. One of the necessary conditions for $\theta(a)(\Gamma)(\brck\phi)$ to satisfy the Truth Lemma is that $$max\{r\in\rats_n\mid L^a_r\phi\in\Gamma\}\leq\theta(a)(\Gamma)(\brck\phi)<min\{r\in\rats_n\mid \lnot L^a_r\phi\in\Gamma\}.$$

Let $\Omega[\psi]$ be the set of $\lng[\psi]$-maximally consistent sets of formulas. $\Omega[\psi]$ is finite and any $\Lambda\in\Omega[\psi]$ contains finitely many formulas modulo propositional equivalence; for the rest of this construction we only count formulas modulo propositional equivalence and we use $\bigwedge\Lambda$ to denote the conjunction of the nontrivial formulas of $\Lambda$.

Ideally would be to construct $\M_\psi$ with $supp(\M_\psi)=\Omega[\psi]$, but this cannot be done because for some $\Lambda\in\Omega[\psi]$, $\{r\in\rats_n\mid \lnot L^a_r\phi\in\Lambda\}$ might be empty. Indeed, while due to (A1) we know that $\{r\in\rats_n\mid L^a_r\phi\in\Lambda\}\neq\emptyset$, the axiomatic system did not provide us with a proof for $\{r\in\rats_n\mid \lnot L^a_r\phi\in\Lambda\}\neq\emptyset$. In fact, such a situation is possible and if it happens, it prevents us to prove the Truth Lemma. A solution could be to convey that $min\emptyset=\infty$. We chose not to do that because we want to have an effective construction that will also involve completing the sets $\Lambda$ with formulas of type $\lnot L^a_r\phi$ when such formulas are absent from $\Lambda$ - this is important latter when we will consider the robustness theorems.

In what follows we construct for each $\Lambda\in\Omega[\psi]$ an extension $\Lambda^+\supseteq[\Lambda]_n$, possibly containing some formulas from $\lng(\A)\setminus\lng[\psi]$, such that for any $\phi\in\Lambda$ and any $a\in\A$, there exists $\lnot L^a_r\phi\in\Lambda^+$. 

We fix an arbitrary $\Lambda\in\Omega[\psi]$ and let $\{\phi_1,...,\phi_i\}\subseteq\Lambda$ be its set of formulas (modulo propositional equivalence). Observe that $\{\phi_1,...,\phi_i\}\vdash\Lambda$. 


\textbf{The construction step [$\phi_1$ versus $\Lambda$]:} 

\noindent From Lemma \ref{construction1} we know that there exists $r\in\prats$ such that $\{\phi_1,...,\phi_i\}\cup\{\lnot L^a_r\phi_1\}$ is consistent. Using the converse of (A2), we obtain that there exists $r\in\rats_n$ such that $\{\phi_1,...,\phi_i\}\cup\{\lnot L^a_r\phi_1\}$ is consistent. This implies that there exists $r\in\mathbb Q_n$ such that $[\Lambda]_n\cup\{\lnot L^a_r\phi_1\}$ is consistent. From Lemma \ref{construction1} we know that there exists $sup\{r\in\prats\mid \{\phi_1,...,\phi_i\}\vdash L^a_r\phi_1\}$, implying that there exists $max\{s\in\mathbb Q_n\mid L^a_s\phi_1\in[\Lambda]_n\}$. Hence, the following values are well defined.
$$y^a_1=min\{s\in\mathbb Q_n\mid [\Lambda]_n\cup\{\lnot L^a_s\phi_1\}\mbox{ is consistent}\},$$ $$x^a_1=max\{s\in\mathbb Q_n\mid L^a_s\phi_1\in[\Lambda]_n\}.$$ About these values we can prove the following lemma.

\begin{lem}\label{construction2}
There exists $r\in\mathbb Q\setminus\mathbb Q_n$ such that $x^a_1<r<y^a_1$ and $\{\lnot L^a_r\phi_1\}\cup [\Lambda]_n$ is consistent.
\end{lem}

\proof
Suppose otherwise, then $\vdash \bigwedge\Lambda\to L^a_r\phi_1$ for all $r<y^a_1$ and applying (R2) we get $\vdash\bigwedge\Lambda\to L^a_{y^a_1}\phi_1$. This contradicts the consistency of $[\Lambda]_n\cup\{\lnot L^a_{y^a_1}\phi_1\}$. Obviously, $r\not\in\mathbb Q_n$. 
\qed

With these results in hand, we return to our construction.
Let $n_1=gran\{1/n,r\}$, where $r\in\mathbb Q\setminus\mathbb Q_n$ is (one of) the value(s) mentioned in Lemma \ref{construction2}.
Let $s^a_1=min\{s\in\mathbb Q_{n_1}\mid[\Lambda]_{n_1}\cup\{\lnot L^a_s\phi_1\}\mbox{ is consistent}\}$, $\Lambda^a_1=\Lambda\cup\{\lnot L^a_{s^a_1}\phi_1\}$ and $\Lambda_1=\displaystyle\bigcup_{a\in A}\Lambda^a_1$. 

In this way we have identified a multiple $n_1\in\nat$ of $n$ and constructed a consistent set $\Lambda_1\supseteq\Lambda$ with the property that for any $a\in A$, $\{r\in\rats_{n_1}\mid L^a_r\phi_1\in\Lambda_1\}$ and $\{r\in\rats_{n_1}\mid \lnot L^a_r\phi_1\in\Lambda_1\}$ are both nonempty.


\textbf{The construction step [$\phi_2$ versus $\Lambda_1$]:} 

\noindent We repeat the construction done before, this time for $\phi_2$ and $\Lambda_1$ and we define $$y^a_2=min\{s\in\mathbb Q_{n_1}\mid[\Lambda_1]_{n_1}\cup\{\lnot L^a_s\phi_2\}\mbox{ is consistent}\},$$ $$x^a_2=max\{s\in\mathbb Q_{n_1}\mid L^a_s\phi_2\in[\Lambda_1]_{n_1}\}.$$ As before, there exists $r\in\mathbb Q\setminus\mathbb Q_{n_1}$ such that $x^a_2<r<y^a_2$ and $\{\lnot L^a_r\phi_2\}\cup [\Lambda_1]_{n_1}$ is consistent. Let $n_2=gran\{1/n_1,r\}$.
Let $s^a_2=min\{s\in\mathbb Q_{n_2}\mid[\Lambda]_{n_2}\cup\{\lnot L^a_s\phi_2\}\mbox{ is consistent}\}$, $\Lambda^a_2=\Lambda_1\cup\{\lnot L^a_{s^a_2}\phi_2\}$ and $\Lambda_2=\displaystyle\bigcup_{a\in A}\Lambda^a_2$. 

As a result of this second step of the construction, we have identified a multiple $n_2\in\nat$ of $n_1$ (hence, of $n$) and constructed a consistent set $\Lambda_2\supseteq\Lambda_1\supseteq\Lambda$ with the property that for any $a\in A$, $\{r\in\rats_{n_2}\mid L^a_r\phi_2\in\Lambda_2\}$ and $\{r\in\rats_{n_2}\mid \lnot L^a_r\phi_2\in\Lambda_2\}$ are both nonempty. Moreover, $\Lambda_2$ inherited from $\Lambda_1$ the property that for any $a\in A$, $\{r\in\rats_{n_2}\mid L^a_r\phi_1\in\Lambda_2\}$ and $\{r\in\rats_{n_2}\mid \lnot L^a_r\phi_1\in\Lambda_2\}$ are nonempty, since $\Lambda_1\subseteq\Lambda_2$ and $n_2$ is a multiple of $n_1$.


\textbf{The complete construction:} 

\noindent We repeat the construction step for [$\phi_3$ versus $\Lambda_2$],..,[$\phi_i$ versus $\Lambda_{i-1}$] and in a finite number of steps we will eventually obtain $\Lambda\subseteq\Lambda_1\subseteq...\subseteq\Lambda_i,$ where $\Lambda_i$ is a consistent set containing a finite set of nontrivial formulas. 

As a result of this entire construction, we have identified a multiple $n_i\in\nat$ of $n$ and constructed a consistent set $\Lambda_i\supseteq\Lambda$ with the property that for any $a\in A$ and any nontrivial formula $\phi\in\Lambda$, $\{r\in\rats_{n_i}\mid L^a_r\phi\in\Lambda_i\}$ and $\{r\in\rats_{n_i}\mid \lnot L^a_r\phi\in\Lambda_i\}$ are nonempty. Using Rule (R1) we can extend this result and claim that for any $a\in A$ and any $\phi\in\Lambda$, $\{r\in\rats_{n_i}\mid L^a_r\phi\in\Lambda_i\}$ and $\{r\in\rats_{n_i}\mid \lnot L^a_r\phi\in\Lambda_i\}$ are nonempty. Hereafter, we use $n_\Lambda$ to denote $n_i$. 

We repeat the construction for all $\Lambda\in\Omega[\psi]$. Let $p=gran\{1/n_\Lambda : \Lambda\in\Omega[\psi]\}$, consequently $p$ is a multiple of $n_\Lambda$. Let $\Lambda^+=[\Lambda_i]_p$ and $\Omega^+[\psi]=\{\Lambda^+: \Lambda\in\Omega[\psi]\}$. 

\begin{rem} Any consistent formula $\phi\in\lng[\psi]$ is an element of a set $\Lambda^+\in\Omega^+[\psi]$. For each $\Lambda\in\Omega[\psi]$, each $\phi\in\Lambda$ and each $a\in\A$, there exist $s,t\in\mathbb Q_p$, $s<t$, such that $L^a_s\phi,\lnot L^a_t\phi\in\Lambda^+$. Moreover, for any $\Lambda^+$ there exists a formula $\rho$ such that $\phi\in\Lambda^+$ iff $\vdash\rho\to\phi$; $\rho$ is, for instance, the conjunction of $\bigwedge\Lambda$ and all the extra formulas added to $\Lambda$ during our construction -- the result is however a finite conjunction. 
\end{rem}

Let $\Omega_p$ be the set of $\lng_p(A)$-maximally consistent sets of formulas. We fix an injective function $f:\Omega^+[\psi]\to\Omega_p$ such that for any $\Lambda^+\in\Omega^+[\psi]$, $\Lambda^+\subseteq f(\Lambda^+)$. The existence of this function is guaranteed by the Lindembaum's Lemma.
We denote by $\Omega_p[\psi]=f(\Omega^+[\psi])$. For $\phi\in\lng[\psi]$, let $\llbracket\phi\rrbracket=\{\Gamma\in\Omega_p[\psi]: \phi\in\Gamma\}$. 

With this construction we have accomplished the first step of the finite model construction: $\Omega_p[\psi]$, which is a set of $\lng_p(A)$-maximally consistent sets, will be the support of $\M_\psi$. Observe that since $\Omega_p[\psi]$ is finite, $(\Omega_p[\psi],2^{\Omega_p[\psi]})$ is an analytic space. It remains to define the transition function $\theta$. The challenge here consists in the fact that we need to define $\theta$ such that for any $a\in\A$, $\theta(a)$ is a measurable function between $(\Omega_p[\psi],2^{\Omega_p[\psi]})$ and $\Delta(\Omega_p[\psi],2^{\Omega_p[\psi]})$ and for any $\Gamma\in\Omega_p[\psi]$, $\theta(a)(\Gamma)$ is a measure on $(\Omega_p[\psi],2^{\Omega_p[\psi]})$. Moreover, the model will have to eventually satisfy the Truth Lemma.

\begin{lem}\label{c}\hfill
\begin{enumerate}[\em(1)]
\item $\Omega_p[\psi]$ is finite.
\item $2^{\Omega_p[\psi]}=\{\llbracket\phi\rrbracket : \phi\in\lng[\psi]\}.$
\item For any $\phi_1,\phi_2\in\lng[\psi]$, $\vdash\phi_1\to\phi_2$ iff $\llbracket\phi_1\rrbracket\subseteq\llbracket\phi_2\rrbracket$.
\item For any $\Gamma\in\Omega_p[\psi]$, $\phi\in\lng[\psi]$ and $a\in A$ there exist $$x=max\{r\in\mathbb Q_p\mid L^a_r\phi\in\Gamma\}\mbox{ and }y=min\{r\in\mathbb Q_p\mid \lnot L^a_r\phi\in\Gamma\}.$$ Moreover, $y=x+1/p$.
\end{enumerate}
\end{lem}

\proof 
1 and 2 are trivial consequences of the construction and 3 is a classical result about maximally consistent sets (ultrafilters of Boolean algebras).

4. If $L^a_x\phi,\lnot L^a_y\phi\in\Gamma$, then $x\neq y$. If $x>y$, $L^a_{x}\phi\in\Gamma$ entails (Axiom (A2)) $L^a_{y}\phi\in \Gamma$, contradicting the consistency of $\Gamma$. If $x+1/p<y$, then $L^a_{x+1/p}\phi\not\in\Gamma$, i.e. $\lnot L^a_{x+1/p}\phi\in\Gamma$ implying that $x+1/p\geq y$ - contradiction.
\qed 

The previous Lemma is a good indicator for where the values of $\theta(a)(\Gamma)(\brck\phi)$ should be placed in intervals of granularity $p$, but this is not enough to guarantee the additivity of $\theta(a)(\Gamma)$. For this, we need to be more precise.

Let $\Omega$ be the set of $\lng(\mathcal A)$-maximally consistent sets of formulas. As before, we fix an injective function $g:\Omega_p\to\Omega$ such that for any $\Gamma\in\Omega_p$, $\Gamma\subseteq g(\Gamma)$; we denote $g(\Gamma)$ by $\Gamma^\infty$. 

\begin{lem}\label{cc}
For any $\Gamma\in\Omega_p[\psi]$, $\phi\in\lng[\psi]$ and $a\in A$, there exists $$z=sup\{r\in\mathbb Q\mid L^a_r\phi\in\Gamma^\infty\}=inf\{r\in\mathbb Q\mid\lnot L^a_r\phi\in\Gamma^\infty\}\mbox{ and }x\leq z<y,$$ where $x$ and $y$ are the values defined in Lemma \ref{c}.
\end{lem}

\proof 
Let $x^\infty=sup\{r\in\mathbb Q\mid L^a_r\phi\in\Gamma^\infty\}$ and $y^\infty=inf\{r\in\mathbb Q\mid\lnot L^a_r\phi\in\Gamma^\infty\}$. 

Suppose that $x^\infty<y^\infty$. Then there exists $r\in\mathbb Q$ such that $x^\infty< r<y^\infty$. From the definition of $x^\infty$ and $y^\infty$ we obtain $\lnot L^a_r\phi,L^a_r\phi\in\Gamma^\infty$ - impossible since $\Gamma^\infty$ is consistent. 

Suppose that $x^\infty>y^\infty$. Then there exists $r\in\mathbb Q$ such that $x^\infty>r>y^\infty$. As $\Gamma^\infty$ is maximally consistent we have either $L^a_r\phi\in\Gamma^\infty$ or $\lnot L^a_r\phi\in\Gamma^\infty$. The first case contradicts the definition of $x^\infty$ while the second the definition of $y^\infty$.

Hence, $x\leq z\leq y$. If $z=y$, then $\lnot L^a_z\phi\in\Gamma$ from the definition of $y$. Moreover, since $z=sup\{r\in\mathbb Q\mid L^a_r\phi\in\Gamma^\infty\}$, there exists a sequence $(z_k)_{k\in\nat}\in\prats$ such that $\lm{k}z_k=z$ and $L^a_{z_k}\phi\in\Gamma^\infty$. Applying (A2), we obtain that for any $z'<z$, $L^a_{z'}\phi\in\Gamma^\infty$. Now (R2) proves that $L^a_z\phi\in\Gamma^\infty$. Consequently, we proved that $\lnot L^a_z\phi,L^a_z\phi\in\Gamma^\infty$ which contradicts the consistency of $\Gamma^\infty$.
\qed 

In what follows we denote the value $z$ defined in Lemma \ref{cc} by $a_\phi^\Gamma$. Now we are ready to define $\M_\psi$. 

Let $\theta_\psi:\A\to[\Omega_p[\psi]\to\Delta(\Omega_p[\psi],2^{\Omega_p[\psi]})]$ be defined, for arbitrary $a\in\A$, $\Gamma\in\Omega_p[\psi]$ and $\phi\in\lng[\psi]$, by $$\theta_\psi(a)(\Gamma)(\llbracket\phi\rrbracket)=a_\phi^\Gamma.$$

\begin{lem}\label{measure}
With the previous notation, for arbitrary $a\in\A$, $\Gamma\in\Omega_p[\psi]$ and $\phi\in\lng[\psi]$, $$\theta(a)(\Gamma)\in \Delta(\Omega_p[\psi],2^{\Omega_p[\psi]}).$$
\end{lem}

\proof 
We prove that the function $\theta_\psi(a)(\Gamma):2^{\Omega_p[\psi]}\to\mathbb R^+$ is well defined and a measure on $(\Omega_p[\psi],2^{\Omega_p[\psi]})$. 

Suppose that for $\phi_1,\phi_2\in\lng[\psi]$ we have $\llbracket \phi_1\rrbracket=\llbracket\phi_2\rrbracket$. Then, from Lemma \ref{c}, $\vdash\phi_1\tto\phi_2$ and $\vdash L^a_r\phi_1\tto L^a_r\phi_2$. Hence, $a_{\phi_1}^\Gamma=a_{\phi_2}^\Gamma$ proving that $\theta_\psi(a)(\Gamma)$ is well defined.

Now we prove that $\theta_\psi(a)(\Gamma)$ is a measure. For this we use Lemma \ref{semiring} and we relay on the fact that $\{\llbracket\phi\rrbracket\mid\phi\in\lng[\psi]\}$ is a finite field. Lemma \ref{semiring} guarantees that it is sufficient to prove that $\theta_\psi(a)(\Gamma)(\emptyset)=0$ and that $\theta_\psi(a)(\Gamma)$ is finitely additive, since the continuity from above in 0 for a finite field derives from the monotonicity guaranteed by the finite additivity.

For showing $\theta_\psi(a)(\Gamma)(\emptyset)=0$, we show that for any $r>0$, $\vdash\lnot L_r^a\bot$. This is sufficient, as (A1) guarantees that $\vdash L_0^a\bot$ and $\llbracket\bot\rrbracket=\emptyset$. Suppose that there exists $r>0$ such that $L_r^a\bot$ is consistent. Let $\epsilon\in(0,r)\cap\mathbb Q$. Then (A2) gives $\vdash L_r^a\bot\to L^a_\epsilon\bot$. Hence, $\vdash L_r^a\bot\to (L_r^a(\bot\land\bot)\land L_\epsilon^a(\bot\land\lnot\bot))$ and applying (A3), $\vdash L_r^a\bot\to L_{r+\epsilon}^a\bot$. Repeating this argument, we can prove that $\vdash L_r^a\bot\to L_s^a\bot$ for any $s$ and (R3) proves the inconsistency of $L_r^a\bot$. 

We show now that if $A,B\in 2^{\Omega_p[\psi]}$ with $A\cap B=\emptyset$, then $\theta_\psi(a)(\Gamma)(A)+\theta_\psi(a)(\Gamma)(B)=\theta_\psi(a)(\Gamma)(A\cup B)$. Using Lemma \ref{c}. 2, we can assume that $A=\llbracket\phi_1\rrbracket$, $B=\llbracket\phi_2\rrbracket$ with $\phi_1,\phi_2\in\lng[\psi]$ and $\vdash\phi_1\to\lnot\phi_2$. Let $x_1=\theta_\psi(a)(\Gamma)(A)$, $x_2=\theta_\psi(a)(\Gamma)(B)$ and $x=\theta_\psi(a)(\Gamma)(A\cup B)$. We prove that $x_1+x_2=x$.

Suppose that $x_1+x_2<x$. Then, there exist $\epsilon_1,\epsilon_2\in\mathbb Q^+$ such that $x_1'+x_2'<x$, where $x_i'=x_i+\epsilon_i$ for $i=1,2$. From the definition of $x_i$, $\lnot L^a_{x_i'}\phi_i\in\Gamma^\infty$. Further, using (A4), we obtain $\lnot L^a_{x_1'+x_2'}(\phi_1\lor\phi_2)\in\Gamma^\infty$, implying that $x_1'+x_2'\geq x$ - contradiction.

Suppose that $x_1+x_2>x$. Then, there exist $\epsilon_1,\epsilon_2\in\mathbb Q^+$ such that $x_1''+x_2''>x$, where $x_i''=x_i-\epsilon_i$ for $i=1,2$. But the definition of $x_i$ implies that $L^a_{x_i''}\phi_i\in\Gamma^\infty$. Further, (A3) gives $L^a_{x_1''+x_2''}(\phi_1\lor\phi_2)\in\Gamma^\infty$, i.e. $x_1''+x_2''\leq x$ - contradiction.

Since the sigma-algebra $2^{\Omega_p[\psi]}$ is finite, the previous results are sufficient to prove that indeed $\theta(a)(\Gamma)\in \Delta(\Omega_p[\psi],2^{\Omega_p[\psi]}).$
\qed 

Because the space $(\Omega_p[\psi],2^{\Omega_p[\psi]})$ has a finite support, the previous lemma already proves that our construction is a CMP.

\begin{thm}\label{MP}
With the previous notations, $\M_\psi=(\Omega_p[\psi], 2^{\Omega_p[\psi]},\theta_\psi)\in\mathfrak M(\A)$.
\end{thm}

\begin{rem}\label{parameter}
Before proceeding with the Truth Lemma, notice that the previous construction is parametric in $p$ and that the choice of $p$ is not unique. It has however a lower bound: $p$ is a multiple of $n$, which is the granularity of $\psi$. We do not focus here on algorithms for computing this parameter, but in \cite{Fagin90} the reader can find an algorithm for computing a similar parameter. Later this parameter will play an essential role in the robustness theorems and for this reason we introduce the following notation: if $\psi$ is a consistent formula and $\M_\psi=(\Omega_p[\psi], 2^{\Omega_p[\psi]},\theta_\psi)\in\mathfrak M(\A)$ is (one of) its finite model(s), then we call $p$ the \emph{parameter of} $\M_\psi$ denoted by $par(\M_\psi)$.
\end{rem}

\begin{lem}[Truth Lemma]\label{c8}
If $\phi\in\lng[\psi]$, then \emph{[}$\M_\psi,\Gamma\models \phi$ iff $\phi\in\Gamma$\emph{]}.
\end{lem}

\proof 
Induction on the structure of $\phi$. 

\textbf{The case $\phi=\top$:} We have always $\M_\psi,\Gamma\models \top$ and $\top\in\Gamma$ since $\Gamma$ is $\lng[\psi]$-maximally consistent.

\textbf{The case $\phi=\phi_1\land\phi_2$:} $\M_\psi,\Gamma\models\phi_1\land\phi_2$ iff [for each $i=1,2$, $\M_\psi,\Gamma\models\phi_i$]. Using the inductive hypothesis, this is equivalent to $\phi_1,\phi_2\in\Gamma$ and since $\phi_1\land\phi_2\in\lng[\psi]$, it is equivalent to $\phi_1\land\phi_2\in\Gamma$.

\textbf{The case $\phi=\lnot\rho$:} $\M_\psi,\Gamma\models\lnot\rho$ is equivalent to $\M_\psi,\Gamma\not\models\rho$ which, applying the inductive hypothesis, is equivalent to $\rho\not\in\Gamma$. Since $\rho\in\lng[\psi]$ and $\Gamma$ is $\lng[\psi]$-maximally consistent, this is equivalent to $\lnot\rho\in\Gamma$.

\textbf{The case $\phi=L^a_r\phi'$:} ($\Longrightarrow$) Suppose that $\M_\psi,\Gamma\models \phi$ and $\phi\not\in\Gamma$. Hence $\lnot\phi\in\Gamma$. Let $y=min\{r\in\mathbb Q_p\mid \lnot L^a_r\phi'\in\Gamma\}$. Then, from $\lnot L^a_r\phi'\in\Gamma$, we obtain $r\geq y$. But $\M_\psi,\Gamma\models L^a_r\phi'$ is equivalent with $\theta_\psi(a)(\Gamma)(\llbracket\phi'\rrbracket)\geq r$, i.e. $a_{\phi'}^\Gamma\geq r$.
On the other hand, from the Rule (R2),  $a_{\phi'}^\Gamma<y$ - contradiction.
\\($\Longleftarrow$) If $L^a_r\phi'\in\Gamma$, then $r\leq a_\phi^\Gamma$ and $r\leq\theta_\psi(a)(\Gamma)(\llbracket\phi\rrbracket)$. Hence, $\M_\psi,\Gamma\models L^a_r\phi$.
\qed 

The Truth Lemma proves the finite model property for our logic.

\begin{thm}[Finite model property]
For any $\lng(\mathcal A)$-consistent formula $\phi$, there exists $\M\in\mathfrak M(\A)$ with finite support of cardinality bound by the structure of $\phi$, and there exists $m\in supp(\M)$ such that $\M,m\models\phi$. 
\end{thm}

\proof
The result derives from the Truth Lemma, since the consistency of $\psi\in\lng[\psi]$ guarantees that there exists a $\lng[\psi]$-maximally consistent set $\Gamma\in\Omega_p[\psi]$ such that $\psi\in\Gamma$. But then, from the truth lemma, $\M_\psi,\Gamma\models\psi$.
\qed 

The finite model property allows us to prove the completeness of the axiomatic system.

\begin{thm}[Weak Completeness]\label{completeness1}
The axiomatic system of $\lng(\A)$ is weak-complete with respect to the Markovian semantics, i.e. if $\models\psi$, then $\vdash\psi$.
\end{thm}

\proof 
We have that [$\models\psi$ implies $\vdash\psi$] is equivalent with [$\not\vdash\psi$ implies $\not\models\psi$], that is equivalent with [the consistency of $\lnot\psi$ implies the existence of a model $(\M,m)\in\mathfrak P(\A)$ for $\lnot\psi$] and this is guaranteed by the finite model property.
\qed


\section{Weak Completeness for $\lng^+(\A)$}\label{axioms1}

In this section we extend the work to $\lng^+(\A)$. In Table \ref{AS1} we present a Hilbert-style axiomatization for $\lng^+(\A)$. 

\begin{table}[!h]
$$
   \begin{array}{ll}
        \mbox{(C1):} & \vdash L^a_0\phi\\
        \mbox{(C2):} & \vdash L^a_{r+s}\phi\to \lnot M^a_r\phi,~s>0\\
        \mbox{(C3):} & \vdash \lnot L^a_r\phi\to M^a_r\phi\\
        \mbox{(C4):} & \vdash \lnot L^a_r(\phi\land\psi)\land \lnot L^a_s(\phi\land\lnot\psi)\to \lnot L^a_{r+s}\phi\\
        \mbox{(C5):} & \vdash \lnot M^a_r(\phi\land\psi)\land \lnot M^a_s(\phi\land\lnot\psi)\to \lnot M^a_{r+s}\phi\\
        \mbox{(T1):} & \mbox{If }\vdash\phi\rightarrow\psi\mbox{ then }\vdash L^a_r\phi\to L^a_r\psi\\
        \mbox{(T2):} & \{L^a_r\phi \mid r<s\}\vdash L^a_s\phi\\
        \mbox{(T3):} & \{M^a_r\phi\mid r>s\}\vdash M^a_s\phi\\
        \mbox{(T4):} & \{L^a_r\phi\mid r>s\}\vdash\bot
   \end{array}
$$\caption{\label{AS1}The axiomatic system of $\lng^+(\A)$}
\end{table}

Notice the differences between these axioms and the axioms in Table \ref{AS} and Table \ref{ASPL}. First of all, Axiom (A2) has to be enforced, in the stochastic case, and it takes the form of the axioms (C2) and (C3). In the probabilistic case, there exist De Morgen dualities between the two modal operators encoded by the rules (B3) and (B4); these two are not sound for the stochastic models. Axiom (A3) has been also enforced by (C5). In addition, we have an extra Archimedean rule for $M^a_r$. 

The concepts of \emph{consistency}, \emph{finite-consistent set}, \emph{maximal} and \emph{maximally-consistent sets} are now used in the context of the new provability relation introduced in Table \ref{AS1}.

We prove below that all the theorems of $\lng(A)$ are also theorems of $\lng^+(\A)$ and we state some theorems of $\lng^+(\A)$ that are central for the weak completeness proof of $\lng^+(\A)$.

\begin{lem}\label{LM}\hfill
\begin{enumerate}[\em(1)]
\item $\vdash M^a_r\phi\to\lnot L^a_{r+s}\phi$, $s>0$,
\item $\vdash\lnot M^a_r\phi\to L^a_r\phi,$ 
\item $\vdash L^a_{r+s}\phi\to L^a_r\phi,$ 
\item $\vdash M^a_r\phi\to M^a_{r+s}\phi$,
\item $\vdash L^a_r(\phi\land\psi)\land L^a_s(\phi\land\lnot\psi)\to L^a_{r+s}\phi,$ 
\item $\vdash M^a_r(\phi\land\psi)\land M^a_s(\phi\land\lnot\psi)\to M^a_{r+s}\phi$,
\item If $\vdash\phi\to\psi$, then $\vdash M^a_r\psi\to M^a_r\phi$.
\end{enumerate}
\end{lem}

\proof
(1) and (2) are the converse of (C2) and (C3) respectively.

\noindent (3). If $s=0$ we have a tautology. Otherwise, using (C2) we have $\vdash L^a_{r+s}\phi\to \lnot M^a_r\phi$ and using (2) we obtain $\vdash L^a_{r+s}\phi\to L^a_r\phi$. (4) can be proved similarly.

\noindent (5). Consider an arbitrary $\e>0$. Using (C2) we obtain 
$$\vdash L^a_r(\phi\land\psi)\land L^a_s(\phi\land\lnot\psi)\to \lnot M^a_{r-\e/2}(\phi\land\psi)\land \lnot M^a_{s-\e/2}(\phi\land\lnot\psi).$$ Now if we apply (C5), we get 
$\vdash L^a_r(\phi\land\psi)\land L^a_s(\phi\land\lnot\psi)\to \lnot M^a_{r+s-\e}\phi$ and applying (2), $\vdash L^a_r(\phi\land\psi)\land L^a_s(\phi\land\lnot\psi)\to L^a_{r+s-\e}\phi.$
Since this result is true for any $\e>0$, applying (T2) we get the result. Similarly can be proved (6).

\noindent (7). Let $\e>0$. Using (T1), $\vdash\phi\to\psi$ implies $\vdash L^a_{r+\e}\phi\to L^a_{r+\e}\psi$ implying $\vdash \lnot L^a_{r+\e}\psi\to \lnot L^a_{r+\e}\phi$. Applying (C3), $\vdash \lnot L^a_{r+\e}\psi\to M^a_{r+\e}\phi$. Applying (1) and (4), we get $\vdash M^a_r\psi\to M^a_{r+\e}\phi$ and since this is true for any $\e>0$, (T3) proves  $\vdash M^a_r\psi\to M^a_r\phi$.
\qed

\begin{thm}[Soundness]
The axiomatic system of $\lng^+(\A)$ is sound for the Markovian semantics, i.e., for any $\phi\in\lng^+(\A)$, if $\vdash\phi$ then $\models\phi$.
\end{thm}

\proof
The proof is similar to the proof of soundness for $\lng(\A)$ without major differences. Here we only prove the soundness of (C3) and (T3). Consider an arbitrary CMP $(\M,m)\in\mathfrak P(\A)$ with $\M=(M,\Sigma,\theta)$.

\textbf{(C3):} Suppose that $\M,m\models\lnot L^a_r\phi$. Then, $\theta(a)(m)(\brck\phi)<r$, hence, $\theta(a)(m)(\brck\phi)\leq r$ that is equivalent to $\M,m\models M^a_r\phi$.

\textbf{(T3):} Suppose that for all $r>s$, $\M,m\models M^a_r\phi$, i.e., for all $r>s$, $\theta(a)(m)(\brck\phi)\leq r$. Using the Archimedean property of rationals, we derive that $\theta(a)(m)(\brck\phi)\leq s$. Hence, $\M,m\models M^a_s\phi$.
\qed


In what follows, we prove the weak-completeness via the finite model property, following a similar construction as for $\lng(\A)$. Due to the similarity between the two constructions, in what follows we only present the arguments and prove the results that differ from the case of $\lng(\A)$. 

The notations introduced in the previous section need as well to be adapted to the signature of $\lng^+(\A)$. 

Consider an arbitrary formula $\phi\in\lng^+(\A)$ and let $R\subseteq\prats$ be the set of $r\in\prats$ such that $r$ is the index of an operator of type $L^a_r$ or $M^a_r$ present in the syntax of $\phi$.
\begin{iteMize}{$\bullet$}
\item The \emph{granularity of} $\phi\in\lng$, denoted by $gr(\phi)$ is defined by $gr(\phi)=gr(R)$
\item The \emph{upper bound of $\phi$}, denoted by $max(\phi)$ is defined by $max(\phi)=max(R)$.
\item The \emph{modal depth of} $\phi$, denoted by $md(\phi)$, is defined inductively by 
$$\begin{array}{ll}
md(\phi)= & \left\{
\begin{array}{ll}
0 & \textrm{if } \phi=\top\\
md(\psi) & \textrm{if } \phi=\lnot\psi\\
max\{md(\psi),md(\psi')\} & \textrm{if }\phi=\psi\land\psi'\\
md(\psi)+1 &\textrm{if } \phi=L^a_r\psi\mbox{ or }\phi=M^a_r\psi
\end{array}\right. \\
\end{array}$$
\item The \emph{actions of} $\phi$ is the set $act(\phi)\subseteq\A$ of indexes $a\in\A$ of the operators of type $L^a_r$ or $M^a_r$ present in the syntax of $\phi$. 
\end{iteMize}

In the rest of this section, for arbitrary $n\in\mathbb N$ and $A\subseteq\A$, let $\lng^+_n(A)$ be the sublanguage of $\lng^+(\A)$ that uses only modal operators $L^a_r$ and $M^a_r$ with $r\in\mathbb Q_n$ and $a\in A$. 

For $\Lambda\subseteq\lng^+(\A)$, let $[\Lambda]_n=\Lambda\cup\{\phi\in\lng^+_n(\A):\Lambda\vdash\phi\}$.


\subsection*{The construction of a finite model for a consistent $\lng^+(\A)$-formula}\hfill

Consider a consistent formula $\psi\in\lng^+(\A)$ with $gr(\psi)=n$ and $act(\psi)=A$. We define $$\lng^+[\psi]=\{\phi\in\lng^+_n(A)\mid max(\phi)\leq max(\psi), md(\phi)\leq md(\psi)\}.$$ Let $\Omega[\psi]$ be the set of all $\lng^+[\psi]$-maximally consistent sets of formulas.  

Consider an arbitrary $\Lambda\in\Omega[\psi]$; we construct, as before, an extension $\Lambda^+\supseteq[\Lambda]_n$, possibly containing some formulas from $\lng^+(\A)\setminus\lng^+[\psi]$, such that for any $\phi\in\Lambda$ and any $a\in\A$, there exists $\lnot L^a_r\phi\in\Lambda^+$. This construction is done exactly as for $\lng(\A)$. Observe that in this case, due to (C2) and (C3), there exists $s\in\prats$ such that $\Lambda^+\ni M^a_s\phi$, and if $max\{r\in\prats\mid L^a_r\phi\}>0$, there also exists $s'\in\prats$ such that $\lnot M^a_{s'}\phi$. We can, in fact, prove the following extension of Lemma \ref{c} (we only state the case similar to the case (4) in Lemma \ref{c}, since the cases (1)-(3) remain true with identical proofs).

\begin{lem}\label{c1}
For any $\Gamma\in\Omega_p[\psi]$, $\phi\in\lng^+[\psi]$ and $a\in A$ there exist $$x=max\{r\in\mathbb Q_p\mid L^a_r\phi\in\Gamma\}\mbox{ and }y=min\{r\in\mathbb Q_p\mid \lnot L^a_r\phi\in\Gamma\},$$ $$v=max\{r\in\mathbb Q_p\mid \lnot M^a_r\phi\in\Gamma\}\mbox{ and }w=min\{r\in\mathbb Q_p\mid M^a_r\phi\in\Gamma\}.$$ Moreover, $y=x+1/p$ and $w=v+1/p$.
\end{lem}

Similarly to Lemma \ref{cc}, for $\lng^+(\A)$ one can prove the following lemma.

\begin{lem}\label{cc1}
For any $\Gamma\in\Omega_p[\psi]$, $\phi\in\lng^+[\psi]$ and $a\in A$, there exists $$z=sup\{r\in\mathbb Q\mid L^a_r\phi\in\Gamma^\infty\}=inf\{r\in\mathbb Q\mid\lnot L^a_r\phi\in\Gamma^\infty\}$$ $$=sup\{r\in\mathbb Q\mid \lnot M^a_r\phi\in\Gamma^\infty\}=inf\{r\in\mathbb Q\mid M^a_r\phi\in\Gamma^\infty\}.$$
Moreover, $x\leq z<y$ and $v<z\leq w$, where $x$, $y$, $v$, $w$ are the values defined in Lemma \ref{c1}.
\end{lem}

As before, we denote this $z$ by $a_\phi^\Gamma$ and we proceed with the definition of the model $\M_\psi$. The next theorem is the correspondent of Theorem \ref{MP} and its proof is similar to the proof of Lemma \ref{measure}.

\begin{thm}\label{MP1}
If $\theta_\psi:\A\to[\Omega_p[\psi]\to\Delta(\Omega_p[\psi],2^{\Omega_p[\psi]})]$ is defined for arbitrary $a\in\A$, $\Gamma\in\Omega_q[\psi]$ and $\phi\in\lng^+[\psi]$ by $\theta_\psi(a)(\Gamma)(\llbracket\phi\rrbracket)=a_\phi^\Gamma$, then $\M_\psi=(\Omega_p[\psi], 2^{\Omega_p[\psi]},\theta_\psi)\in\mathfrak M(\A)$.
\end{thm}

This last result allows us to prove the Truth Lemma also for $\lng^+(\A)$. 

\begin{lem}[Truth Lemma]\label{c9}
If $\phi\in\lng^+[\psi]$, then \emph{[}$\M_\psi,\Gamma\models \phi$ iff $\phi\in\Gamma$\emph{]}.
\end{lem}

\proof
In addition to the proof of Lemma \ref{c8}, we need to prove the case $\phi=M^a_r\phi'$.
\\($\Longrightarrow$) Suppose that $\M_\psi,\Gamma\models \phi$ and $\phi\not\in\Gamma$. Hence $\lnot\phi\in\Gamma$. Let $v=max\{r\in\mathbb Q_p\mid \lnot M^a_r\phi'\in\Gamma\}$. Then, from $\lnot M^a_r\phi'\in\Gamma$, we obtain $r\leq v$. But $\M_\psi,\Gamma\models M^a_r\phi'$ is equivalent with $\theta_\psi(a)(\Gamma)(\llbracket\phi'\rrbracket)\leq r$, i.e. $a_{\phi'}^\Gamma\leq r$.
On the other hand, from Lemma \ref{cc1},  $a_{\phi'}^\Gamma>v$ - contradiction.
\\($\Longleftarrow$) If $M^a_r\phi'\in\Gamma$, then $r\geq a_\phi^\Gamma$ and $r\geq\theta_\psi(a)(\Gamma)(\llbracket\phi\rrbracket)$. Hence, $\M_\psi,\Gamma\models M^a_r\phi$.
\qed

As before, the truth lemma implies the finite model property and the weak completeness theorem for $\lng^+(\A)$ with Markovian semantics.

\begin{thm}[Finite Model Property]
For any $\lng^+(\mathcal A)$-consistent formula $\phi$, there exists $\M\in\mathfrak M(\A)$ with finite support of cardinality bound by the structure of $\phi$, and there exists $m\in supp(\M)$ such that $\M,m\models\phi$. 
\end{thm}

\begin{thm}[Weak Completeness]
The axiomatic system of $\lng^+(\A)$ is weak complete with respect to the Markovian semantics, i.e. if $\models\psi$, then $\vdash\psi$.
\end{thm}

\begin{rem}\label{parameter1}
Before ending this section we shall insist on the fact that the finite model construction for $\lng^+(\A)$, as for $\lng(\A)$, is parametric in $p$ and that the choice of $p$ is not unique. 
As before, if $\psi$ is a consistent formula and $\M_\psi=(\Omega_p[\psi], 2^{\Omega_p[\psi]},\theta_\psi)\in\mathfrak M(\A)$ is (one of) its finite model(s), then we call $p$ the \emph{parameter of} $\M_\psi$ denoted by $par(\M_\psi)$.
\end{rem}


\section{Strong completeness of Markovian Logics and their canonical models}\label{canonic}

In this section we address the problem of strong completeness, both for $\lng(\A)$ and $\lng^+(\A)$. The strong completeness requires to prove that any consistent theory $\Gamma$ is satisfied by some model (process). The Markovian logics with the axiomatizations presented in the previous sections are not strongly complete and in order gain this property one needs to enrich the axiomatic systems with the so-called \emph{Countable Additivity Rule} and to assume the \emph{Lindembaum property} that every consistent set of formulas (in the new axiomatic system) has a maximally consistent extension. The adoption of the Lindembaum property as a postulate rather than a property to be proved was firstly proposed by Goldblatt in \cite{Goldblatt10} where he shows that this choice is unavoidable in order to achieve the strong completeness of such logics for coalgebras over measurable spaces. In the absence of these assumptions, one cannot build a a canonic model from maximally consistent sets of formulas. Instead, the canonical models can only be constructed from \emph{truth sets} (sets satisfied by some process), as done in \cite{Aumann99b,Moss06}. Zhou proves in \cite{Zhou11} that for such logics the class of of truth sets of formulas is a proper subclass of that of maximally consistent sets of formulas.

In this paper we are interested in the strong completeness and in constructing canonical models from maximally consistent sets of formulas that will eventually provide us a useful tool for studying metric properties of the space of logical formulas. To accomplish this, we assume in what follow the Lindembaum property and we consider the concept of provability obtained by enriching the axiomatic systems in Table \ref{AS} and Table \ref{AS1} with the Boolean rules for infinitary deduction and Countable Additivity Rule (CAR) stated below. The results and the constructions presented in this section can be developed in a similar manner for both $\lng(\A)$ and $\lng^+(\A)$ and for this reason in what follows we use $\lng$ to range over the set $\{\lng(\A),\lng^+(\A)\}$ and we present the arguments at this level of generality.

Given an arbitrary set $\Phi\subseteq\lng$ of formulas, let $\bigwedge\Phi$ denote the set of the (finite) conjunctions of the elements of $\Phi$ and for arbitrary $a\in\A$ and $r\in\mathbb Q^+$, let $L^a_r\Phi=\{L^a_r\phi\mid\phi\in\Phi\}$.

$$\mbox{(CAR): For arbitrary $\Phi\subseteq\lng$ and $\phi\in\lng$,}~~~~~~~~\Phi\vdash\phi\mbox{ implies }L^a_r\bigwedge\Phi\vdash L^a_r\phi.$$

For the beginning we shall prove the soundness of (CAR) for the Markovian semantics.

\begin{thm}[Soundness of (CAR)]
$$\mbox{For arbitrary $\Phi\subseteq\lng$ and $\phi\in\lng$,}~~~~~~~~\Phi\models\phi\mbox{ implies }L^a_r\bigwedge\Phi\models L^a_r\phi.$$
\end{thm} 

\proof
Let $\llbracket\Phi\rrbracket$ be the set of models satisfying all the formulas of $\Phi$. 

Consider an arbitrary CMP $(\M,m)\in\mathfrak P(\A)$ with $\M=(M,\Sigma,\theta)$, such that $(\M,m)\models L^a_r\Phi$. This means that for each finite conjunction $\psi$ of elements of $\Phi$ we have that $(\M,m)\models L^a_r\psi$, which is equivalent to $\theta(a)(m)(\llbracket\psi\rrbracket)\geq r$. 

Suppose that $\bigvee\Phi=\{\phi_1,\phi_2,...\}$ and for each $i\in\mathbb N$, let $\psi_i=\phi_1\land...\land\phi_i$. Observe that for each $i$, $\llbracket\psi_i\rrbracket\supseteq\llbracket\psi_{i+1}\rrbracket$ and $\displaystyle\bigcap_{i\in\mathbb N}\llbracket\psi_i\rrbracket=\llbracket\Phi\rrbracket$. Consequently, $\lm{i}\theta(a)(m)(\llbracket\psi_i\rrbracket)=\theta(a)(m)(\llbracket\Phi\rrbracket)$. 

Since for each $i$, $\psi_i\in\Phi$, we have that $\theta(a)(m)(\llbracket\psi_i\rrbracket)\geq r$ implying $\lm{i}\theta(a)(m)(\llbracket\psi_i\rrbracket)\geq r$. Hence, $\theta(a)(m)(\llbracket\Phi\rrbracket)\geq r$.

Further, observe that $\Phi\models\phi$ means that $\llbracket\Phi\rrbracket\subseteq\llbracket\phi\rrbracket$. Hence, $\theta(a)(m)(\llbracket\phi\rrbracket)\geq r$.
\qed

\bigskip

In what follows we concentrate on proving the strong completeness. We reuse some of the notation introduced in the previous sections.

Let $\Omega$ be the set of $\lng$-maximally consistent sets of formulas. For arbitrary $\phi\in\lng$, let 
$\prth{\phi}= \{\Phi\in\Omega\mid \phi\in\Phi\}$ and $\prth{\lng}=\{\prth{\phi}\mid\phi\in\lng\}.$

Similar results with the ones in Lemma \ref{cc} and Lemma \ref{cc1} can be proved for this extended concept of consistency and using them, we can define for arbitrary $\Phi\in\Omega$, $a\in\A$ and $\phi\in\lng$, $$a^{\Phi}_\phi=sup\{r\in\mathbb Q: L^a_r\phi\in\Phi\}=inf\{r\in\mathbb Q:\lnot L^a_r\phi\in\Phi\}=$$ $$inf\{r\in\mathbb Q:M^a_r\phi\in\Phi\}=sup\{r\in\mathbb Q: \lnot M^a_r\phi\in\Phi\}.$$ 
This allows us to prove the existence of the canonical model for $\lng\in\{\lng(\A),\lng^+(\A)\}$. In the next theorem $\prth{\lng}^\sigma$ denotes the sigma-algebra induced by $\prth{\lng}$.

\begin{thm}[Canonical Model]
With the previous notation, $\M_\lng=(\Omega,\prth{\lng}^\sigma,\theta_\lng)\in\mathfrak M(\A)$, where $\theta_\lng:\A\to[\Omega\to\Delta(\Omega,\prth{\lng}^\sigma)]$ is defined for arbitrary $a\in\A$, $\Phi\in\Omega$ and $\phi\in\lng$ by $\theta_\lng(a)(\Phi)(\llbracket\phi\rrbracket)=a_\phi^{\Phi}$.
\end{thm}

\proof
This proof is similar to the proofs of the Theorems \ref{MP} and \ref{MP1}. The only additional things to prove are that $(\Omega,\prth{\lng}^\sigma)$ is an analytic space and the countable-additivity for the functions $\theta_\lng(a)(\Phi)$ required to guarantee that these functions are indeed measures. These results were not necessary for the Theorems \ref{MP} and \ref{MP1} because the supports of the finite models were finite and consequently also their $\sigma$-algebras.

That $(\Omega,\prth{\lng}^\sigma)$ is an analytic space derives from the fact that it is a Polish space, and this can be proved as in \cite{Zhou12}, Corollary 4.5. 
                                                                                                                                                        
From the proof of Lemma \ref{measure}, which can be reproduced identically for our case, we know that for arbitrary $\Phi\in\Omega$ and $a\in\A$, $\theta_\lng(a)(\Phi)$ is finitely additive on $\prth{\lng}$ and $\theta_\lng(a)(\Phi)(\emptyset)=0$ (because $\llbracket\bot\rrbracket=\emptyset$). Since $\prth{\lng}$ is a field, in order to prove that $\theta_\lng(a)(\Phi)$ can be uniquely extended to a metric on $\prth{\lng}^\sigma$, it is sufficient to prove that $\theta_\lng(a)(\Phi)$ is continuous from above in 0 on $\prth{\lng}$, as stated in Lemma \ref{semiring}.

Let $\Psi=\{\psi_1,\psi_2,...\}\subseteq\lng$ be such that for each $i\in\mathbb N$, $\prth\psi_i\supseteq\prth\psi_{i+1}$ and $\displaystyle\bigcap_{i\in\mathbb N}\prth\psi_i=\emptyset$.
\\Let $x_i=\theta_\lng(a)(\Phi)(\prth{\psi_i})$. To prove that $\theta_\lng(a)(\Phi)$ is continuous from above in 0, it is sufficient to prove that $\lm{i}x_i=0$. 

Since $\theta_\lng(a)(\Phi)$ is positive and monotone, there exists $y=\lm{i}x_i$. Suppose that $y\neq 0$; then, there exists a rational $p$ such that $0<p<y$. From here we obtain that there exists a $j$ such that for each $i\geq j$, $x_i>p$; we can assume, without loosing generality, that $j=1$. Hence, $L^a_p\psi_i\in\Phi$ for each $i$.

Observe that since $\prth\psi_i\supseteq\prth\psi_{i+k}$ for each $i,k\in\mathbb N$, $\phi\in\bigwedge\Psi$ implies that there exists $k\in\mathbb N$ such that $\phi=\psi_k$. Consequently, [$L^a_p\psi_i\in\Phi$ for each $i\in\mathbb N$] guarantees that $L^a_p\Psi\subseteq\Phi$. Hence, $L^a_p\Psi$ must be consistent (as a subset of a consistent set). 

Because $p>0$, $\not\vdash L^a_p\bot$ and using (CAR), $\Psi\not\vdash\bot$. Hence, $\psi$ is consistent and using Lindenbaum property, there exists $\Phi'\in\Omega$ such that $\Psi\subseteq\Phi'$. But then $\Phi'\in\displaystyle\bigcap_{i\in\mathbb N}\prth{\psi_i}$, contradicting the hypothesis that $\displaystyle\bigcap_{i\in\mathbb N}\prth\psi_i=\emptyset$.

This concludes the proof that $y=0$ and $\theta_\lng(a)(\Phi)$ is continuous from above in 0 on $\prth\lng$.
\qed

Notice in the previous proof the central role played by (CAR) and Lindenbaum property. In the previous version of this paper \cite{Cardelli11a} we have omitted these aspects.

The existence of the canonical models allows us to prove the most general truth lemma.

\begin{lem}[Extended Truth Lemma]
With the previous notations, for arbitrary $\phi\in\lng$ and $\Phi\in\Omega$, $$\M_\lng,\Phi\models \phi\mbox{ iff }\phi\in\Phi.$$
\end{lem}

\proof
Due to the way we have used the $\lng$-maximally consistent sets to construct the finite model, the proof of this lemma derives from the proofs of the other truth lemmas \ref{c8} and \ref{c9}.
\qed

We conclude this section with the Strong Completeness Theorem that is a straightforward consequence of the previous lemma and the Lindenbaum property.

\begin{thm}[Strong Completeness]
For arbitrary $\phi\in\lng$ and arbitrary consistent set $\Phi\subseteq\lng$, $$\Phi\models\phi\mbox{ iff }\Phi\vdash\phi.$$
\end{thm}


\section{From bisimulation to the metric space of logical formulas}\label{characterize}

One of the main motivation for studying quantitative logics for probabilistic and stochastic processes was, since the first papers on this subject \cite{Larsen91}, the characterization of stochastic/probabilistic bisimilarity. 
To start with, we state that a version of the Hennesy-Milner theorem holds for Markovian logic: the logical equivalences induced by $\lng(\A)$ and by $\lng^+(\A)$ on the class of CMPs coincide with stochastic bisimilarity. The proof follows closely the proof of the corresponding result for probabilistic systems presented in \cite{zigzag,Panangaden09} and for this reason we will not present it here. However, it consists in showing that the negation free-fragment of $\lng(\A)$ characterizes stochastic bisimilarity, while negation and $M^a_r$ do not differentiate bisimilar processes. 

\begin{thm}[Logical characterization of stochastic bisimilarity]\label{modalcharact}\hfill
\\Let $\M=(M,\Sigma,\tau),\M'=(M',\Sigma',\tau')\in\mathfrak M(\A)$, $m\in M$ and $m'\in M'$. The following assertions are equivalent. 
\begin{enumerate}[\em(1)]
\item $(\M,m)\sim(\M',m')$;
\item For any $\phi\in\lng(\A)$, $\M,m\models\phi$ iff $\M',m'\models\phi$;
\item For any $\phi\in\lng^+(\A)$, $\M,m\models\phi$ iff $\M',m'\models\phi$.
\end{enumerate}
\end{thm}

A consequence of the logical characterization of bisimulation is that any CMP is bisimilar to a CMP of the canonical model, as stated in the Representation Theorem stated below. This derives immediately from the logical characterization of bisimilarity and Extended Truth Lemma.

\begin{thm}[Representation Theorem]
Any process $P\in\mathfrak P(\A)$ is bisimilar to some process $(\M_\lng,\Phi)$ of the Canonical Model. More exactly, $P\sim(\M_\lng,\Phi)$ where $\Phi=\{\phi\in\lng\mid P\models\phi\}$.
\end{thm}

\bigskip

The concept of stochastic/probabilistic bisimilarity is however a very strict concept: it only verifies whether two processes have identical behaviours. In applications we need instead to know whether two processes that may differ by a small amount in the real-valued parameters (rates or probabilities) have similar behaviours. To solve this problem a class of pseudometrics have been proposed in the literature \cite{Desharnais04,Panangaden09,vanBreugel01b,vanBreugel03}, to measure how similar two processes are in terms of stochastic/probabilistic behaviour. Because these pseudometrics are quantitative relaxations of the bisimulation relation, they can be defined relying on the Markovian logics. Since in what follows the development works identically both for $\lng(\A)$ and $\lng^+(\A)$, we simply use $\lng$ to denote any of the two.

Formally, for the class $\mathfrak P(\A)$ of Markov processes and for the Markovian logic $\lng$, the behavioural pseudometric can be induced by a function $d:\mathfrak P(\A)\times\lng\to\preals$ which extends the (characteristic function of the) satisfiability relation $\models:\mathfrak P(\A)\times\lng\to\{0,1\}$. The function $d$ quantifies with strictly positive numbers the situations in which a process does not satisfy a property \cite{Desharnais04,Panangaden09}.  

To exemplify, in this paper we work with the function $d:\mathfrak P(\A)\times\lng\to\preals$, defined below for the set $\mathfrak P(\A)$ of CMPs and $\lng\in\{\lng^+(\A),\lng(\A)\}$.

$d((\M,m),\top)=0$,

$d((\M,m),\lnot\phi)=1-d((\M,m),\phi)$,

$d((\M,m),\phi\land\psi)=max\{d((\M,m),\phi),d((\M,m),\psi)\}$,

$d((\M,m),L^a_r\phi)=\langle r,\theta(a)(m)(\llbracket\phi\rrbracket)\rangle$,

$d((\M,m),M^a_r\phi)=\langle\theta(a)(m)(\llbracket\phi\rrbracket),r\rangle$,
\\where for arbitrary $r,s\in\mathbb R_+$, 
$$\begin{array}{ll}
\langle r,s\rangle= & \left\{
\begin{array}{ll}
r-s & \textrm{if } r-s>0\\
0 & \textrm{otherwise}
\end{array}\right. \\
\end{array}$$
The results presented in this section relay on the fact that $d$, as most of the functions that quantify satisfiability for stochastic or probabilistic logics, is defined on top of the transition function $\theta$. For this reason, these results can be similarly proved for other bisimulation pseudometrics. 

The first result states that $d$ characterizes stochastic bisimilarity.

\begin{lem}\label{metric}
If $(\M,m),(\M',m')\in\mathfrak P(\A)$, then $$(\M,m)\sim(\M',m')\mbox{ iff } \emph{[} \forall\phi\in\lng, d((\M,m),\phi)=d((\M',m'),\phi)\emph{]}.$$
\end{lem}

\proof 
($\Longrightarrow$) Induction on $\phi$. The Boolean cases are trivial and the cases $\phi=L^a_r\psi$ and $\phi=M^a_r\psi$ derive from the fact that $\theta(a)(m)(\llbracket\psi\rrbracket)=\theta'(a)(m')(\llbracket\psi\rrbracket)$. 
\\($\Longleftarrow$)  For an arbitrary $\phi\in\lng$ and $a\in\A$, we have that for all $r\in\prats$, $$d((\M,m), L^a_r\phi)=d((\M',m'),L^a_r\phi);$$ and for $r$ big enough $$d((\M,m), L^a_r\phi)=r-\theta(a)(m)(\llbracket\phi\rrbracket)\enspace,\quad d((\M',m'), L^a_r\phi)=r-\theta'(a)(m')(\llbracket\phi\rrbracket).$$ Hence, $\theta(a)(m)(\llbracket\phi\rrbracket)=\theta'(a)(m')(\llbracket\phi\rrbracket)$ which implies $(\M,m)\sim(\M',m')$.
\qed 

As we have anticipated, a function $d:\mathfrak P(\A)\times\lng\to\preals$ which characterizes bisimulation in the sense of Lemma \ref{metric}, induces a distance between stochastic processes, $$D:\mathfrak P(\A)\times\mathfrak P(\A)\to\preals$$ defined for arbitrary $P,P'\in\mathfrak P(\A)$ by 
$$D(P,P')=sup\{|d(P,\phi)-d(P',\phi)|,\phi\in\lng\}.$$
The next lemma states that $D$ is a pseudometric and its kernel is the stochastic bisimilarity.

\begin{lem}
The function $D:\mathfrak P(\A)\times\mathfrak P(\A)\to\preals$ defined before is a pseudometric such that $$D(P,P')=0\mbox{ iff }P\sim P'.$$
\end{lem}

\proof
Obviously $D$ is symmetric. We prove the triangle inequality. For arbitrary $P,Q,R\in\mathfrak P(\A)$ and $\phi\in\lng$, we have the following inequalities 
$$|d(P,\phi)-d(Q,\phi)|\leq |d(P,\phi)-d(R,\phi)|+|d(R,\phi)-d(Q,\phi)|,$$
$$\supr{\phi\in\lng}|d(P,\phi)-d(Q,\phi)|\leq \supr{\phi\in\lng}(|d(P,\phi)-d(R,\phi)|+|d(R,\phi)-d(Q,\phi)|),$$
$$\supr{\phi\in\lng}|d(P,\phi)-d(Q,\phi)|\leq \supr{\phi\in\lng}|d(P,\phi)-d(R,\phi)|+\supr{\phi\in\lng}|d(R,\phi)-d(Q,\phi)|.$$
Hence, $D(P,Q)\leq D(P,R)+D(R,Q)$.

Now we prove that the kernel of $D$ is bisimulation.
\\If $P\sim P'$, Lemma \ref{metric} guarantees that for any $\phi\in\lng$, $d(P,\phi)=d(P',\phi)$ implying that $\supr{\phi\in\lng}|d(P,\phi)-d(P',\phi)|=0$. 
\\Reverse, if $\supr{\phi\in\lng}|d(P,\phi)-d(P',\phi)|=0$, then for any $\phi\in\lng$, $d(P,\phi)=d(P',\phi)$ and Lemma \ref{metric} guarantees that $P\sim P'$.\qed

Similarly, one can use $d$ to define a pseudometric $\ol d:\lng\times\lng\to[0,1]$ over the space of logical formulas by 
$$\ol d(\phi,\psi)=sup\{|d(P,\phi)-d(P,\psi)|, P\in\mathfrak P(\A)\},\mbox{ for arbitrary }\phi,\psi\in\lng.$$

\begin{lem}\label{metric1}
The function $\ol d:\lng\times\lng\to[0,1]$ defined before is a pseudometric and $$\ol d(\phi,\psi)=\ol d(\lnot\phi,\lnot\psi).$$
\end{lem}

\proof 
We prove that it satisfies the triangle inequality. We have $$\supr{P\in\mathfrak P(\A)}|d(P,\phi)-d(P,\psi)|+\supr{P\in\mathfrak P(\A)}|d(P,\psi)-d(P,\rho)|\geq$$ 
$$\supr{P\in\mathfrak P(\A)}(|d(P,\phi)-d(P,\psi)|+|d(P,\psi)-d(P,\rho)|)\geq \supr{P\in\mathfrak P(\A)}|d(P,\phi)-d(P,\rho)|.\eqno{\qEd}$$

\noindent This construction allows us to introduce the first robustness theorem stating that the perturbation of a logical property bounds the effect on the function $d$.

\begin{thm}[Strong Robustness]
For arbitrary $\phi,\psi\in\lng$ and $P\in\mathfrak P(\A)$, $$d(P,\psi)\leq d(P,\phi)+\ol d(\phi,\psi).$$
\end{thm}

\proof 
From the definition of $\ol d$ we have that $d(P,\psi)-d(P,\phi)\leq \ol d(\phi,\psi)$.
\qed 

Notice the importance of this result: while the values of $d$ depend on the model, $\ol d$ refers exclusively to logical properties and it is, apparently, independent of the model. The proof of this independence is not trivial and we will not present it here; for this proof the reader is referred to \cite{Larsen12b}. If one can evaluate $\ol d$ independently of the model, then this theorem can be used in many applications where one needs to evaluate the value of $d$ for extremely large systems. Using maybe techniques such as statistical model checking, one can get an evaluation of $d(P,\phi)$ and further use our robustness theorem to get boundaries for $d(P,\psi)$ for arbitrary $\psi\in\lng$.

Similar constructions can be done for any class of stochastic or probabilistic models for which a correspondent logic that characterizes bisimulation has been defined. But despite the obvious utility of the robustness theorem, in most of the cases such a result is not computable due to the definition of $\ol d$ that involves the quantification over the entire class of continuous Markov processes.

This is exactly where the weak and strong complete axiomatizations of $\lng(\A)$ and $\lng^+(\A)$ and the finite model properties play their role. In what follows, we use the construction of the finite model for an $\lng$-consistent formula presented in the previous sections to effectively compute an approximation of $\ol d$ within a given error $\varepsilon>0$. Below we reuse the notations of the finite model constructions introduced in Section \ref{axioms} and Section \ref{axioms1}. 

The next lemma states that $\ol d$ can be characterized using the processes of the canonical model $\M_\lng$. In this way, it implicitly relates $\ol d$ to provability, since these processes are $\lng$-maximally consistent sets of formulas. 

\begin{lem}\label{metric2}
With the previously used notations, for arbitrary $\phi,\psi\in\lng$, $$\ol d(\phi,\psi)=\supr{\Phi\in\Omega}|d((\M_\lng,\Phi),\phi)-d((\M_\lng,\Phi),\psi)|.$$
\end{lem}

\proof 
Any $(\M,m)\in\mathfrak M(\A)$ satisfies a maximally-consistent set of formulas, hence there exists $\Phi\in\Omega$ such that $(\M,m)\sim(\M_\lng,\Phi)$, i.e., for any $\phi\in\lng$, $d((\M,m),\phi)=d((\M_\lng,\Phi),\phi)$. 
\qed

In what follows we reduce the quantification space in the characterization of $\ol d$ presented in the previous lemma from $\Omega$ to the domain of a finite model. Central for this is the construction of the finite model for a consistent formula. 

For an arbitrary consistent formula $\psi\in\lng$, let $\M_\psi=(\Omega_p[\psi], 2^{\Omega_p[\psi]},\theta_\psi)\in\mathfrak M(\A)$ be the model of $\psi$ constructed in the previous section; and let $p=par(\M_\psi)$ be the parameter of $\M_\psi$ discussed in Remark \ref{parameter} and Remark \ref{parameter1}. 

Let $\s{d}:\lng[\psi]\times\lng[\psi]\to\preals$ be defined, for arbitrary $\phi,\phi'\in\lng[\psi]$, as follows.
$$\s{d}(\phi,\phi')= \displaystyle\max_{\Gamma\in\Omega_p[\psi]}|d((\M_\psi,\Gamma),\phi)-d((\M_\psi,\Gamma),\phi')|.$$
 Notice that $\s d$ is similar to $\ol d$ except that it takes the supremum over a finite set of processes, hence, a maximum. There is however a strong relation between $\ol d$ and $\s d$ involving the parameter $par(\M_\psi)$, as described in the next lemma. 
 
\begin{lem}\label{metrics3}
With the notations previously introduced, for arbitrary $\phi,\phi'\in\lng[\psi]$, $$\ol d(\phi,\phi')\leq\s{d}(\phi,\phi')+2/p.$$
\end{lem}

\proof 
To prove the inequality, we prove first that for arbitrary $\phi\in\lng[\psi]$, $$|d((\M_\lng,\Gamma^\infty),\phi)-d((\M_\psi,\Gamma),\phi)|\leq 1/p,$$ where we reuse the notation of the finite model construction.
We do the proof by induction on $\phi$.

\textbf{The case $\phi=\top$:} $d((\M_\lng,\Gamma^\infty),\top)=d((\M_\psi,\Gamma),\top)=0$.

\textbf{The case $\phi=\lnot\phi'$:} $|d((\M_\lng,\Gamma^\infty),\phi)-d((\M_\psi,\Gamma),\phi)|=|1-d((\M_\lng,\Gamma^\infty),\psi)-1+d((\M_\psi,\Gamma),\psi)|=|d((\M_\lng,\Gamma^\infty),\psi)-d((\M_\psi,\Gamma),\psi)|$. And now we can apply the inductive hypothesis.

\textbf{The case $\phi=\phi'\land\phi''$:} $|d((\M_\lng,\Gamma^\infty),\phi'\land\phi'')-d((\M_\psi,\Gamma),\phi'\land\phi'')|=$ $$|max\{d((\M_\lng,\Gamma^\infty),\phi'),d((\M_\lng,\Gamma^\infty),\phi'')\}-max\{d((\M_\psi,\Gamma),\phi'),d((\M_\psi,\Gamma),\phi'')\}|\leq 1/p,$$
since $|d((\M_\lng,\Gamma^\infty),\phi')-d((\M_\psi,\Gamma),\phi')|\leq 1/p$ and $|d((\M_\lng,\Gamma^\infty),\phi'')-d((\M_\psi,\Gamma),\phi'')|\leq 1/p$ from the inductive hypothesis.

\textbf{The case $\phi=L^a_r\phi'$:} Since $\Gamma\subseteq\Gamma^\infty$, from the way we constructed $\M_\psi$ we obtain that $$|\theta_\psi(a)(\Gamma)(\brck{\phi'})-\theta_\lng(a)(\Gamma^\infty)(\brck{\phi'})|\leq1/p.$$
$\bullet$ If $r\leq
  \theta_\lng(a)(\Gamma^\infty)(\brck{\phi'})<\theta_\psi(a)(\Gamma)(\brck{\phi'})$,
  then $d((\M_\lng,\Gamma^\infty),L^a_r\phi')=d((\M_\psi,\Gamma),L^a_r\phi')=0$, hence, $|d((\M_\lng,\Gamma^\infty),L^a_r\phi')-d((\M_\psi,\Gamma),L^a_r\phi')|=0$.
\\{\sloppy $\bullet$ If $
  \theta_\lng(a)(\Gamma^\infty)(\brck{\phi'})<r\leq\theta_\psi(a)(\Gamma)(\brck{\phi'})$,
  then we have  $d((\M_\psi,\Gamma),L^a_r\phi')=0$ and $d((\M_\lng,\Gamma^\infty),L^a_r\phi')=r-\theta_\lng(a)(\Gamma^\infty)(\brck{\phi'})$, hence, $|d((\M_\lng,\Gamma^\infty),L^a_r\phi')-d((\M_\psi,\Gamma),L^a_r\phi')|=r-\theta_\lng(a)(\Gamma^\infty)(\brck{\phi'})\leq\theta_\psi(a)(\Gamma)(\brck{\phi'})-\theta_\lng(a)(\Gamma^\infty)(\brck{\phi'})\leq 1/p$.}
\\$\bullet$ If $ \theta_\lng(a)(\Gamma^\infty)(\brck{\phi'})<\theta_\psi(a)(\Gamma)(\brck{\phi'})\leq r$, then $d((\M_\psi,\Gamma),L^a_r\phi')=r-\theta_\psi(a)(\Gamma)(\brck{\phi'})$ and $d((\M_\lng,\Gamma^\infty),L^a_r\phi')=r-\theta_\lng(a)(\Gamma^\infty)(\brck{\phi'})$, hence, $|d((\M_\lng,\Gamma^\infty),L^a_r\phi')-d((\M_\psi,\Gamma),L^a_r\phi')|=|(r-\theta_\lng(a)(\Gamma^\infty)(\brck{\phi'}))-(r-\theta_\psi(a)(\Gamma)(\brck{\phi'}))=|\theta_\psi(a)(\Gamma)(\brck{\phi'})-\theta_\lng(a)(\Gamma^\infty)(\brck{\phi'})|\leq 1/p$.
\\$\bullet$ All the other sub cases are treated similarly.

\textbf{The case $\phi=M^a_r\phi'$:} can be proved similarly to the case $\phi=L^a_r\phi'$.

Now we prove our inequality. Since $\s{d}(\phi,\phi')= \displaystyle\max_{\Gamma\in\Omega_p[\psi]}|d((\M_\psi,\Gamma),\phi)-d((\M_\psi,\Gamma),\phi')|$, there exists $\Gamma_0\in\Omega_p[\psi]$ that realizes this maximum.

Using the inequality that we just proved, we obtain that for arbitrary $\Gamma\in\Omega_p[\psi]$, $$|d((\M_\lng,\Gamma^\infty),\phi)-d((\M_\lng,\Gamma^\infty),\phi')|\leq|d((\M_\psi,\Gamma),\phi)-d((\M_\psi,\Gamma),\phi')|+2/p,$$ and since $\Gamma_0$ realizes the maximum, $$|d((\M_\lng,\Gamma^\infty),\phi)-d((\M_\lng,\Gamma^\infty),\phi')|\leq|d((\M_\psi,\Gamma_0),\phi)-d((\M_\psi,\Gamma_0),\phi')|+2/p.$$ Because this inequality is satisfied by any $\Gamma$, we can take it to the limit and obtain $$\supr{\Gamma^\infty\in\Omega}|d((\M_\lng,\Gamma^\infty),\phi)-d((\M_\lng,\Gamma^\infty),\phi')|\leq|d((\M_\psi,\Gamma_0),\phi)-d((\M_\psi,\Gamma_0),\phi')|+2/p,$$ i.e., $\ol d(\phi,\phi')\leq\s{d}(\phi,\phi')+2/p.$

\qed

This last result finally allows us to prove a weaker version of the robustness theorem which evaluates $d((\M,m),\psi)$ from $d((\M,m),\phi)$, based on $\s{d}(\phi,\psi)$ and a given error.

\begin{thm}[Weak Robustness]
For an arbitrary consistent formula $\psi\in\lng$, arbitrary $\phi,\phi'\in\lng[\psi]$ and arbitrary $P\in\mathfrak P(\A)$, $$d(P,\phi)\leq d(P,\phi')+\s{d}(\phi,\phi')+2/p,$$ where $p=par(\M_\psi)$.
\end{thm}

\proof
The result derives from combining the Strong Robustness Theorem with Lemma \ref{metrics3}.

\qed

The Weak Robustness Theorem is a very useful tool to estimate $d(P,\phi)$ given $d(P,\phi')$ within a given error $2/p$. With respect to the Strong Robustness Theorem, it guarantees that the evaluation can be done in finite time, since $\s d$ requires to investigate a finite CMP. The finite model construction can be used to provide an algorithm for computing the parameter $p$. A similar algorithm for probabilistic case can be find in \cite{Fagin90}.


\section{Conclusions and related works}

We have introduced Continuous Markovian Logic, a multimodal logic designed to specify quantitative and qualitative properties of continuous Markov processes. CML is endowed with operators that approximate the rates of the labelled transitions of CMPs. This logic characterizes the stochastic bisimilarity of CMPs.

We have presented weak and strong complete axiomatizations for the entire logic as well as for its fragment without $M^a_r$-operators. These axiomatic systems are significantly different from the probabilistic case and from each other. The weak completeness proofs relay on the finite model properties. The constructions of the finite models adapts the filtration method of modal logics to stochastic settings, where a series of specific problems had to be solved. The finite model constructions and the complete axiomatizations allow us to approach the problems of approximating bisimulation-distances and to prove a series of new results such as the robustness theorems.

This paper opens a series of interesting research questions regarding the relationship between satisfiability, provability and metric semantics. There are many open questions related to the definition of $\ol d$ and to the structure of the metric space of formulas. One of the problems, that we postpone for future work, is finding a classification of the functions $d$ to reflect properties of $\ol d$. For instance, in collaboration with Prakash Panangaden \cite{Larsen12b}, we have proven that if $d$ satisfies some specific continuity conditions, then $\ol d$ characterizes the logical equivalence between formulas without negation in the probabilistic case and between formulas without negation and without $M^a_r$ operators in the stochastic case, i.e., only for some special $\phi$ and $\psi$ we have [$\ol d(\phi,\psi)=0$ iff $\vdash\phi\tto\psi$]. However, the problem of finding a formal characterization of the kernel of $\ol d$ remains open. In the same paper we gave topological characterizations of the pseudometric space of logical formulas for probabilistic and for the Markovian logics. At the topological level one can see better the differences between the two logics. For the probabilistic logic the situation is quite simple: the set $\brck\phi$ of models satisfying $\phi$ is closed whenever $\phi$ does not involve negation and open otherwise. For the Markovian logics the situation is much more complicated and the sets $\brck\phi$ can, in various cases, produce open, closed, $G_{\delta}$ or $F_{\sigma}$ sets of models\footnote{In topology, a $G_{\delta}$ set is a countable intersection of open sets and a $F_{\sigma}$ set in a countable union of closed sets.}. 

There exist, however, distances enjoying even stronger properties such as [$\models\phi\to\psi$ iff $\forall P\in\mathfrak P(\A)$, $d(P,\psi)\leq d(P,\phi)$]. Each of these metrics organizes the set of logical formulas as a pseudometric space with specific topological properties. The complete axiomatization is probably the key for classifying such structures and for understanding the relationship between the topological space of models and the topological space of logical formulas.


\section*{Acknowledgement}
This research was supported by the VKR Center of Excellence MT-LAB and by the Sino-Danish Basic Research Center IDEA4CPS. Mardare was also supported by Sapere Aude: DFF-Young Researchers Grant 10-085054 of the Danish Council for Independent Research. 

Mardare would like to thank to Prakash Panangaden, Dexter Kozen, Chunlai Zhou and Ernst-Erich Doberkat for discussions regarding various aspects of the theory of Markov processes and logics that eventually allowed us to arrive to the current level of understanding of these problems. 

We are also thankful to the anonymous referees of our paper that helped us with their comments and suggestions. 





\end{document}